\documentclass[final,authoryear,3p,times]{elsarticle}

\newcommand{\volume}{\mathop{\ooalign{\hfil$V$\hfil\cr\kern0.08em--\hfil\cr}}\nolimits}

\usepackage{natbib}
\usepackage{amssymb}
\usepackage{graphicx}
\usepackage{verbatim}
\usepackage[mathscr]{eucal}
\usepackage{epstopdf} 
\newcommand{\dtilde}[1]{\tilde{\tilde{#1}}}

\usepackage{multirow} 
\usepackage{amsmath}  
\usepackage{float}    
\usepackage{setspace}

\journal{Ocean Modelling, \normalfont{10.1016/j.ocemod.2020.101705}}

\begin{document}
\begin{frontmatter}

\author{Bjarke Eltard Larsen\corref{cor1}\fnref{DTU}}
\ead{bjelt@mek.dtu.dk}
\cortext[cor1]{Corresponding author}
\author[Aberdeen]{Dominic A. van der A}
\author[Twente]{Joep van der Zanden}
\author[Utrecht]{Gerben Ruessink}
\author[DTU]{David R. Fuhrman}


\title{Stabilized RANS simulation of surf zone kinematics and boundary layer processes beneath large-scale plunging waves over a breaker bar}

\address[DTU]{Technical University of Denmark, Department of Mechanical Engineering, DK-2800 Kgs. Lyngby, Denmark}
\address[Aberdeen]{School of Engineering, University of Aberdeen, Aberdeen, AB24 3UE, United Kingdom}
\address[Twente]{Offshore department, Maritime Research Instititute Netherlands (MARIN),, Haagsteeg 2
6708 PM Wageningen, Netherlands}
\address[Utrecht]{Department of Physical Geography, Faculty of Geosciences, Institute
for Marine and Atmospheric Research Utrecht, Utrecht University, Utrecht,
Netherlands}

\begin{abstract}
This paper presents numerical simulations of a bichromatic wave group propagating and breaking over a fixed breaker bar. The simulations are performed using a newly stabilized Reynolds-averaged Navier Stokes (RANS) two-equation turbulence closure, which solves the longstanding problem of over-production of turbulence beneath surface waves in the nearly potential flow region prior to breaking. This model has previously been tested on small-scale spilling breaking regular waves, whereas in this work focus is on full (rather than model) scale application, wave groups (rather than regular waves) and plunging (rather than spilling) breakers. Additionally this paper has novel emphasis on bottom boundary layer dynamics which are very important for cross-shore sediment transport predictions.  
The model is validated by comparing with results from a previous experimental campaign. The model is shown to predict the surface elevations, velocities and turbulence well in the shoaling and outer surf-zone, avoiding turbulence over-production and incorrect undertow structure typical of standard turbulence closures. Comparison with detailed boundary layer measurements in the shoaling position reveals that the model is able to accurately capture the temporal dynamics of the entire wave boundary layer, including evolution of the boundary layer thickness, velocity overshoot and phase-shifts. Comparison in the surf zone additionally reveals that the model is able to accurately capture the transport of breaking-induced turbulence into the wave boundary layer. The performance of the model indicates that it can be used directly in the simulation of cross-shore sediment transport and morphology and also be used to study important hydrodynamic processes, which can help improve the predictive skill of morphodynamic profile models applied in coastal engineering.
\end{abstract}
\end{frontmatter}

\section{Introduction}
While long-shore sediment transport can be predicted with reasonable accuracy, net cross-shore sediment transport rates are extremely difficult to predict due to the variety of hydrodynamic processes that are involved (e.g. undertow, various forms of streaming, asymmetry and skewness wave shape effects, as well as wave breaking-induced turbulence). These processes lead to suspended and bed load transport components that are often comparable in magnitude, but which differ in terms of direction (i.e.~onshore vs.~offshore). 
Only limited studies have been performed with detailed measurements of flow and turbulence within the boundary layer in the surf zone (\citealp[see the large scale plunging wave experiments by][]{vanderZandenetal2016,vanderZandenetal2018a,Fromantetal2019,vanderZandenetal2019}\citealp[, the large scale experiments by] []{Andersonetal2017,Mierasetal2017,Mierasetal2019} \citealp[ or the small scale breaking wave experiments by][]{Coxetal1996,Henriquezetal2014}). These studies have provided valuable new insights such as demonstrating the transport of wave breaking turbulence into the boundary layer, showing that advection rather than diffusion carries breaking-induced turbulence into the wave boundary layer and illustrating that turbulence in the boundary layer in the surf zone can sustain over multiple wave cycles. Despite these findings, turbulence and wave boundary layer dynamics in the shoaling and surf zone are still not fully understood and some of these processes have yet to be incorporated into predictive empirical models for coastal engineering practise. Furthermore, experiments such as the ones mentioned above, are time consuming and they typically suffer from a relatively coarse spatial resolution.

Computational fluid dynamics (CFD), either through a Reynolds-averaged Navier Stokes (RANS) approach, using a Large Eddy Simulation (LES) approach \citep[see e.g.][]{Christensen2006,Zhouetal2017} or a Smoothed Particle Hydrodynamics (SPH) approach \citep[see e.g.][]{Loweetal2019} can potentially handle the breaking processes and boundary layer dynamics naturally. CFD simulations of breaking waves are still computationally demanding and it is often necessary to work with the least computationally demanding of the three CFD approaches, namely the RANS models.    
Most past research using RANS models, studying processes relevant for cross-shore sediment transport, have focused only on the boundary layer dynamics using one-dimensional-vertical (1DV) models \citep[see e.g.][]{Fuhrmanetal2009a,Fuhrmanetal2009b,Ruessinketal2009,Kranenburgetal2012} or only on the outer flow surf zone kinematics with limited focus on resolving the wave bottom boundary layer \citep[see e.g.][]{LinLiu1998,Bradford2000,Chellaetal2015,LupieriContero2015,Brown2016,Derakhtietal2016a,Derakhtietal2016b,Devolderetal2018}. There are only a few papers that have used RANS models to simulate surf-zone kinematics, including the free-surface, and have also focused on near-bed hydrodynamic processes and sediment transport.  \cite{Jacobsenetal2014} and \cite{JacobsenFredsoe2014} presented a fully-coupled hydrodynamic and morphological model (called \verb|sediMorph|), and used it to simulate breaker bar development. The same model was used by \cite{FernandezMoraetal2016} to simulate the mobile bed experiments of  \cite{vanderZandenetal2016}. In this case morphology was turned off, and the focus was purely on sediment transport. More recently, \cite{Kimetal2018} used the so-called \verb|sedWaveFOAM| model  to simulate sheet flow under non-breaking waves and \cite{Kimetal2019} further showed SedWaveFoam works for the sheet flow under breaking waves on the crest of a sandbar.
These models have shown great potential, but fundamentally rely on the ability to accurately simulate the kinematics (including the boundary layer) in the shoaling and surf zone. 

Using RANS models to simulate shoaling and surf zone processes is not trivial. Past studies have shown a tendency to significantly overestimate turbulence levels in simulations of breaking waves, and this has even been most pronounced prior to breaking.
This problem was originally diagnosed  by \cite{MayerMadsen2000} and recently \cite{LarsenFuhrman2018} showed that seemingly all (at least all they analyzed) widely used RANS models (both $k$-$\epsilon$ and $k$-$\omega$ type models) are unconditionally unstable in nearly potential flow (characterized by a low rotation rate) regions beneath waves, resulting in non-physical exponential growth of the turbulent kinetic energy (TKE) and eddy viscosity. They demonstrated how such models can be formally and easily stabilized, by modifying an already established stress-limiting feature within the eddy viscosity formulation.  Using a stabilized model, \cite{LarsenFuhrman2018} showed significant improvements in predicted turbulence levels and undertow profiles, especially prior to breaking and in the outer surf zone. 

In the study by \cite{LarsenFuhrman2018} the stabilized model was only tested on small-scale simulations of regular waves propagating and breaking (spilling) on a fixed slope. It is therefore still uncertain how the model performs in large-scale conditions, with more complex bathymetries, non-regular wave conditions and other breaking types. In the study of \cite{LarsenFuhrman2018} there was likewise no focus on the near-bed hydrodynamics, which is of utmost importance for simulating sediment transport.
This paper therefore aims to investigate the ability of stabilized RANS models to simulate hydrodynamic processes relevant for cross-shore sediment transport, such as spatio-temporal turbulence variations and boundary layer dynamics.
This will be done by simulating a recent large-scale experiment \citep{vanderZandenetal2019}, that involves bichromatic waves propagating and breaking over a fixed bar. Using previously unpublished boundary layer data from this experiment, complemented by results from the model, this paper further aims to improve existing knowledge on near-bed flow and turbulence  beneath surface waves in both the shoaling region and the surf zone. 

The present paper extends the work of \cite{LarsenFuhrman2018}, focusing on full rather than model scale application, wave groups rather than regular waves, plunging rather than spilling breakers and with novel emphasis on boundary layer dynamics. The latter is especially important since a good understanding of the hydrodynamic performance in the boundary layer is necessary before attempting to simulate sediment transport processes.  

This paper is organized as follows. The experimental setup and model description are given in Section \ref{sec:Setup}. Section \ref{sec:Results} presents and compares experimental and numerical results for the outer flow (i.e. well above the boundary layer) with a focus on water surface elevations in Section \ref{sec:surface}, outer flow velocities in Section  \ref{sec:outer} and turbulence in Section \ref{sec:outerk}. Section \ref{sec:NearBed} presents and compares experimental and numerical results in the boundary layer with a focus on near bed velocities in Section \ref{sec:BL} and turbulence in Section \ref{sec:BLk}. Section \ref{sec:Discussion} contains a broad discussion on both models' performance, the implications for cross-shore sediment transport as well as a discussion on RANS/VOF models capabilities in simulating breaking waves. Finally, conclusions are presented in Section \ref{sec:Conclusion}.

\section{Experimental set-up and model description}
\label{sec:Setup}

\subsection{Experimental set-up}
As the experimental setup and data treatment have already been presented in detail in \cite{vanderZandenetal2019}, only a summary will be presented here. 
The experiments were conducted in the 100 m long, 3 m wide and 5 m deep wave flume at the Polytechnic University in Barcelona. In the flume a breaker bar, that was initially formed during a previous mobile-bed experiment using monochromatic plunging waves \citep{vanderZandenetal2016}, was fixed with a top layer of concrete.  The same fixed breaker bar set-up was used previously to study monochromatic breaking waves \citep{vanderAetal2017,vanderZandenetal2018a}. For the present experiments, large roughness elements with median grain diameter $d_{50}=9$ mm were glued onto the concrete (from $x >35$ m) to create a larger boundary layer, with the aim of enabling detailed measurements within the (relatively large) wave boundary layer.

\begin{figure*}
  \centering
    \includegraphics[trim=0.5cm 3cm 0cm 0cm, clip=true, scale=1]{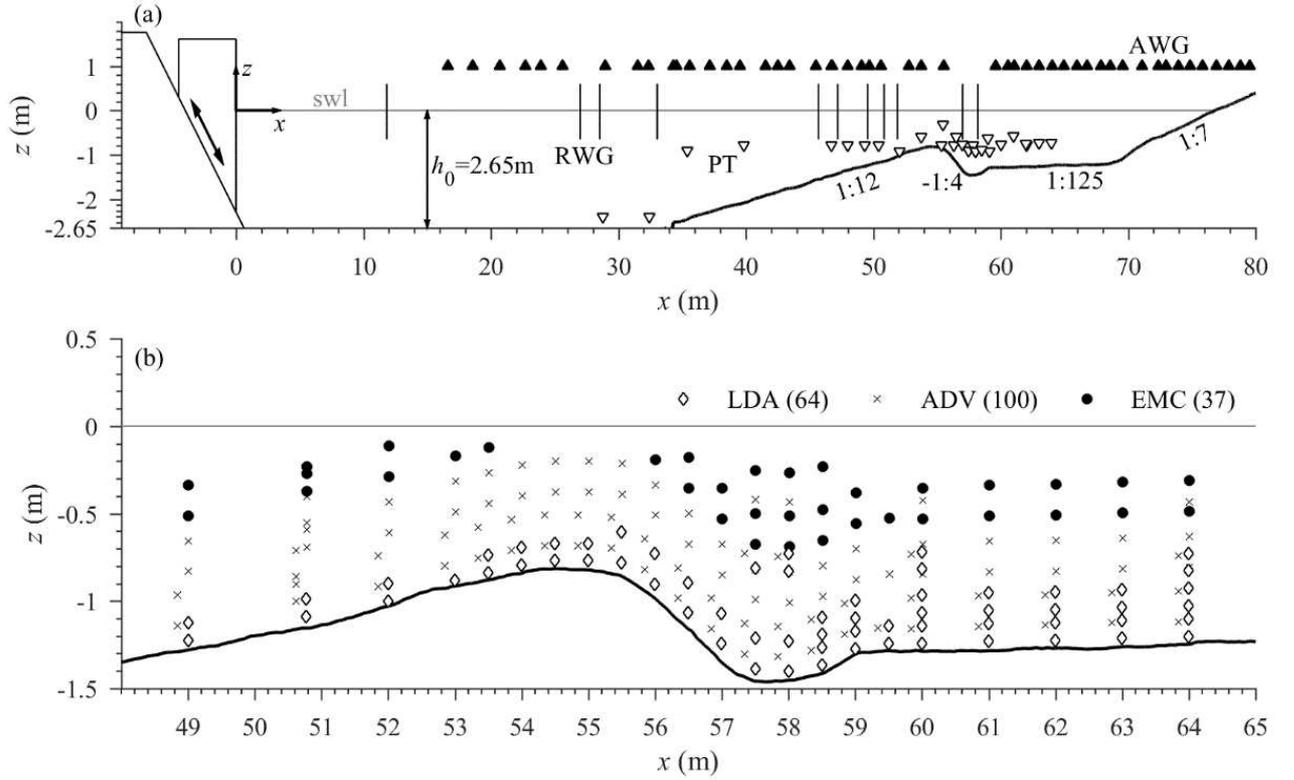}
\caption{Experimental set-up including measurement locations by: (a) resistive wave gauges (RWG, solid vertical lines), pressure transducers (PT, white triangles), and acoustic wave gauges (AWG, black triangles); (b) laser Doppler anemometers (LDA, diamonds), acoustic Doppler velocimeters (ADVs, crosses), and electromagnetic current meters (EMC, circles), with numbers between brackets indicating the total number of measurement locations by each instrument.  Figure adopted from \cite{vanderZandenetal2019} }   
    \label{fig:setup}
 \end{figure*}

Figure \ref{fig:setup} shows a sketch of the setup, together with instrument positions. 
Water surface elevations were measured at 92 cross-shore locations using a combination 
of resistive and acoustic wave gauges as well as pressure transducers, deployed from the side-walls of the flume. Pressure measurements were converted to surface elevations using the approach for non-linear waves described in \cite{Bonnetonetal2018}. Flow velocity was measured at 201 unique cross-shore and vertical positions using acoustic, optic, and electromagnetic current meters, deployed from a mobile frame that could be moved along the length and height of the flume (Figure \ref{fig:setup}). 

The incoming waves were generated using first order theory and consisted of a bichromatic wave group which had period components $T_1 = 3.9375$ s and $T_2 = 4.5$ s, resulting in a wave group with a group period $T_{gr} = 1/(f_1-f_2)$ = 31.5 s, where $f_1$ and $f_2$ are the two frequency components of the group. (Note that this definition of the wave group is different from the one based on the full cycle of the wave envelope as e.g. presented in \cite{FredsoeDeigaard1992}). Each group consisted of 7.5 short waves with mean period $T_m = 4.2$ s. The measured maximum wave height for the flat part of the flume was $H_{max}\approx0.58$--$0.64$ m.
A ramp-up period of approximately 10 groups was used to ensure stable conditions in the flume, and subsequent data from approximately 100 groups was used for phase-averaging. TKE and turbulent Reynolds stresses were quantified from the measurements through a Reynolds decomposition. A high spatial measurement coverage was obtained by realizing 48 repeats of the experiment, with the mobile frame each time positioned at different cross-shore locations and elevations.
In the experiments $z_{bed}$ was defined as the top of the roughness elements over an area of 0.3 x 0.3 m$^2$ with an estimated uncertainty of $\mathcal{O}$(5 mm) ($ \approx 0.5k_s$, $k_s$ being Nikuradse's equivalent sand roughness).

The TKE of the experiments was for the ADVs calculated as 

\begin{equation}
k = \frac{1}{2} \left(\overline{u^{\prime^2}} + \overline{v^{\prime^2}}+ \overline{w^{\prime^2}}\right).
\end{equation}
The LDA only measured two components of the velocity ($u, w$) and TKE was derived from the phase-averaged measurements following \cite{Svendsen1987}

\begin{equation}
k = \frac{2}{3} \left(\overline{u^{\prime^2}} + \overline{w^{\prime^2}}\right).
\end{equation}
For more details on the setup and data treatment, see \cite{vanderZandenetal2019}.

\subsection{Model description}

The simulations are performed using the two-phase volume-of-fluid method (VOF) flow model \verb|waves2FOAM| developed by \cite{Jacobsenetal2012} and implemented in \verb|foam-extend-3.1|. In this model, the Reynolds-averaged Navier-Stokes (RANS) \eqref{eq:RANS} and continuity \eqref{eq:cont} equations are solved:

\begin{equation}
\frac{\partial (\rho u_i)}{\partial t} +  \frac{\partial (\rho u_i u_j)}{\partial x_j} = -\frac{\partial p^*}{\partial x_i} -g_j x_j \frac{\partial \rho}{\partial x_i} + \frac{\partial}{\partial x_j} \left( 2 S_{ij} (\mu+\mu_T)  \right),
\label{eq:RANS}
\end{equation}

\begin{equation}
\frac{\partial u_i}{\partial x_i} = 0.
\label{eq:cont}
\end{equation}
Here $u_i$ are the ensemble averaged components of the velocities, $x_i$ are the
Cartesian coordinates,  $\rho$ is the density, $t$ is time, $S_{ij}$
is the mean strain rate tensor given by

\begin{equation}
S_{ij} = \frac{1}{2} \left( \frac{\partial u_i}{\partial x_j} + \frac{\partial u_j}{\partial x_i}\right),
\end{equation}
$\mu=\rho\nu$ is the dynamic molecular viscosity, $\nu$ is the kinematic viscosity and $\mu_T=\rho\nu_T$ is the dynamic eddy viscosity. The latter originates from the Reynolds stress tensor, $\tau_{ij}$, which is expressed
according to the Boussinesq approximation

\begin{equation}
\tau_{ij} = -\overline{u_i^{'} u_j^{'}} = 2 \nu_T S_{ij} - \frac{2}{3} k \delta_{ij}.
\label{eq:Boussinesq}
\end{equation}
Here the overbar signifies time (ensemble) averaging, $\nu_T$ is the kinematic eddy
viscosity, $\delta_{ij}$ is the Kronecker delta, and 

\begin{equation}
k = \frac{1}{2}\overline{u_i^{'}u_i^{'}}
\label{eq:k}
\end{equation}
is the turbulent kinetic energy density.  In these equations a prime
superscript denotes turbulent (fluctuating) velocity components.
In contrast to normal presentation of the RANS equations the turbulent normal stresses are not included in the last term in \eqref{eq:RANS} (which now only contains $\mu_T$), but instead are included in 
$p^*=p_e+2/3 \rho k \delta_{ij}$ which is a modified mean pressure containing the pressure in excess of hydrostatic pressure $p_e$, and the turbulent normal stresses, $2/3 \rho k \delta_{ij}$, as described in \cite{Pope2000}, p. 88.

To close the system the stabilized $k$-$\omega$ model described in \cite{LarsenFuhrman2018} is used, which is a generalization of the \cite{Wilcox2006} model. This model solves transport equations for $k$

\begin{equation}
\frac{\partial (\rho k)}{\partial t} +  \frac{\partial (u_j \rho k)}{\partial x_j} = 
\rho P_k 
-\rho P_b
-  \rho \beta^{*}k\omega + \frac{\partial}{\partial x_j} \left[\left(\mu +  \rho \sigma^{*} \frac{k}{\omega}\right)\frac{\partial k}{\partial x_j}\right] 
\label{eq:keq}
\end{equation}
and the specific dissipation rate $\omega$:

\begin{equation}
\begin{split}
\frac{\partial (\rho \omega)}{\partial t} +\frac{\partial (u_j \rho \omega)}{\partial x_j} =  \rho P_\omega - \rho  \beta \omega^2 +  \rho \frac{\sigma_d}{\omega}\frac{\partial k}{\partial x_j} \frac{\partial \omega}{\partial x_j} + \\
\frac{\partial}{\partial x_j} \left[\left(\mu +  \rho \sigma \frac{k}{\omega}\right)\frac{\partial \omega}{\partial x_j}\right].
\end{split}
\label{eq:omegaeq}
\end{equation}
The shear production term for $k$ is 

\begin{equation}
P_k=\tau_{ij} \frac{\partial u_i}{\partial x_j} = p_0\nu_T,
\hspace{1cm}
p_0= 2 S_{ij}S_{ij}.
\label{eq:Pk}
\end{equation}
Similarly, the buoyancy production for $k$ is formulated as

\begin{equation}
P_b = -\frac{g_i}{\rho} \overline{\rho' u_i'}  = p_b\nu_T,
\hspace{0.5cm} p_b=\alpha_b^* N^2,
\hspace{0.5cm} N^2 = \frac{g_i}{\rho}\frac{\partial\rho}{\partial x_i},
\label{eq:Pb}
\end{equation}
where $(g_1,g_2,g_3)=(0,0,-g)$ is gravitational acceleration and $N^2$ is the square
of the Brunt-Vaisala frequency.  The production of $\omega$ is
likewise taken as

\begin{equation}
P_\omega =
\alpha \frac{\omega}{k}\frac{\tilde{\omega}}{\dtilde{\omega}} P_k
= \alpha\frac{\omega}{\dtilde{\omega}}p_0.
\label{eq:Pomega}
\end{equation}
In this model the eddy viscosity is defined as 

\begin{equation}
\nu_T = \frac{k}{\tilde{\omega}}
\label{eq:nuT}
\end{equation}
with

\begin{equation}
\dtilde{\omega} =
\mbox{max}\left[ \omega , \lambda_1 \sqrt{\frac{p_0-p_b}{\beta^{*}}} \right],
\label{eq:omegadTilde}
\end{equation}
\begin{equation}
\tilde{\omega}=\mbox{max}\left[\dtilde{\omega}, \lambda_2 \frac{\beta}{\beta^{*} \alpha} \frac{p_0}{p_{\Omega}} \omega\right].
\label{eq:wtildeNew}
\end{equation}

The standard closure coefficients utilized are those of \cite{Wilcox2006}: $\alpha = 0.52$, $\beta=0.0708$ (constant for 2D flows),
$\beta^{*}=0.09$, $\sigma=0.5$, $\sigma^{*}=0.6$, $\sigma_{do}=0.125$, with

\begin{equation}
\sigma_d = H \left( \frac{\partial k}{\partial x_j} \frac{\partial \omega}{\partial x_j} \right) \sigma_{do},
\end{equation}
where $H \left(\cdot\right)$ is the Heaviside step function, which
takes a value of unity if the argument is positive and zero otherwise.
Additionally, we adopt the value $\alpha_b^*=1.36$, which was derived by \cite{LarsenFuhrman2018}, as well as the stress limiting coefficients  $\lambda_1=0.2$ and $\lambda_2=0.05$. It is worth noting that the formally stabilized model can be turned into a standard (still buoyancy modified) model by setting $\lambda_2=0$. Some results with this variant will be shown in this paper to illustrate the difference between standard and stabilized closures. It is emphasized that the stabilization suggested by in \cite{LarsenFuhrman2018} is not limited to the free-surface region, but rather includes the entire nearly potential flow region beneath waves. For a thorough discussion of this see \cite{FuhrmanLarsen2020}.

A scalar field $\gamma$ is  used to track the two fluids, where 
$\gamma=0$ represents pure air and $\gamma=1$ pure water, with any intermediate value representing a mixture.  The distribution of $\gamma$ is governed by the advection equation

\begin{equation}
\frac{\partial \gamma}{\partial t} + \frac{\partial u_j \gamma}{\partial x_j} + \frac{ \partial  u^{r}_j \gamma (1-\gamma)}{\partial x_j} = 0,
\end{equation}
where $u_j^r$ is a relative velocity used to compress the interface. The method is
developed by OpenCFD$^\text{\textregistered}$, and is documented in \cite{Berberovicetal2009}. The density and dynamic viscosity of any cell are calculated as

\begin{equation}
\rho = \gamma \rho_{water} + (1-\gamma)\rho_{air},
\end{equation}

\begin{equation}
\mu = \gamma \mu_{water} + (1-\gamma)\mu_{air}.
\end{equation}

The bottom has a no-slip condition imposed such that velocities at the bed are zero. 
The friction velocity $U_f$ is determined from the
tangential velocity at the nearest cell center based on an assumed
rough logarithmic velocity distribution

\begin{equation}
\frac{u}{{U_f}}
=
\frac{1}{\kappa}\ln\frac{30 z_c}{k_s},
\label{eqn:ulog}
\end{equation} 
where $\kappa=0.4$ is the von Karman constant, $k_s$ is again
Nikuradse's equivalent sand grain roughness, and $z_c=\Delta z/2$ is
the normal distance from the theoretical wall to the nearest cell center, with
$\Delta z$ being the cell thickness next to the wall. The friction
velocity is then used to calculate $k$ and $\omega$ in the cell
nearest to the wall with standard wall functions, here given in
both dimensional and dimensionless forms:

\begin{equation}
k = \frac{U_f^2}{\sqrt{\beta^*}}
\hspace{1cm}\mbox{or}\hspace{1cm}
\frac{k}{U_f^2} = \frac{1}{\sqrt{\beta^*}},
\label{eqn:kwall}
\end{equation}
\begin{equation}
\omega = \frac{U_f}{\sqrt{\beta^*}\kappa z_c}
\hspace{1cm}\mbox{or}\hspace{1cm}
\frac{\omega\nu}{U_f^2}=\frac{1}{\sqrt{\beta^*}\kappa z_c^+},
\label{eqn:omegawall}
\end{equation}
where $z_c^+=z_c U_f/\nu$ is the distance to the near-wall
cell center in wall coordinates. At the wall, to maintain consistency with the numerically calculated velocity gradients, the eddy viscosity is calculated from

\begin{equation}
U_f^2= \frac{|\tau_b|}{\rho} =\left(\nu+\nu_T\right) \frac{\partial |u_t|}{\partial n},
\end{equation}
as described by \cite{SumerFuhrman2020}. Here $u_t$ is the tangential velocity  and $n$ is the direction normal to the wall.

In the present simulations  $k_s=1.4 d_{50} = 0.0126$ m is used rather than the conventional $k_s=2.5 d_{50}$ as this is more in line with the findings from \cite{Fuhrmanetal2010} with a $d=7$ mm bed  and \cite{SchlichtingGersten2003} with a $d=26$ mm bed, and this value also provides reasonable results compared to the recent measurements as will be shown. 

The simulations are performed in two dimensions. The bottom in the simulations follows the measured barred profile of the experiment, using a running average over 20 cm in the horizontal to smooth out small bed variations due to individual grains. The small step seen around $x=34$ m (Figure \ref{fig:setup}a) was smoothed to avoid having to resolve the flow around this step in detail. The computational grid is composed of $2250\times 188$ cells ($x$ and $z$ direction), yielding a total of 423,000 cells. The grid follows the bathymetry with the majority of the cells having  $\Delta x=\Delta z=0.04$ m, corresponding to an aspect ratio of unity \citep[as recommended by][]{Jacobsenetal2012,Roenbyetal2017a,Larsenetal2019}. Near the bed the cells are gradually refined with near-bed cells having $\Delta z_{cell} = k_s/7=0.0018$ m which ensures a high vertical resolution of the wave bottom boundary layer and which is consistent with the rough wall boundary condition which requires cell centres to be positioned above $k_s/30$, see Equation \eqref{eqn:ulog}. It should be noted that in the model the vertical distance is measured from the theoretical wall (positioned approximately $0.25 k_s$ beneath the top of the roughness elements). Therefore to ensure the same bed level in model and experiments, the vertical coordinate in the experiments was increased  by $\Delta z = 0.25 k_s = 3.2$ mm. The uncertainty in the experimental bed location has little importance when comparing outer flow quantities but it may be important when comparing near-bed quantities where rapid vertical variation in both turbulence and velocity exists.  

In the simulations the time step has been adjusted such that a maximum Courant
number $Co =|u_n|\Delta t/\Delta x_n =0.05$ (summation over indices suppressed) is maintained at all times, where
$u_n$ is the normal velocity component at a cell face and $\Delta x_n$ is distance between the two cells centres connected by the face. Such a low $Co$ is not common in the simulation of surface waves, but was shown to greatly increase the accuracy of the predicted wave kinematics in \cite{Larsenetal2019}.

The waves were generated using the unidirectional version of the second order bichromatic bidirectional solution by \cite{MadsenFuhrman2006} with an added return current ensuring zero net mass flux. A relaxation zone was employed at 0 m $\leq x \leq$ 10 m.  The waves in the simulations had $T_1 = 3.9375$s and $T_2 = 4.5$s as in the experiments and $H_1=0.325$ m and $H_2=0.31$ m to produce a maximum wave height similar to that of the experiments in the flat part of the domain. 

Similar to the experiments a ramp-up period was needed to achieve quasi-steady conditions in the flume. In both the model and the experiments 10 groups were used as a ramp-up period. Following this ramp-up period, another 10 groups were generated, and the presented results, in what follows, have been averaged and phase-averaged over these additional 10 groups. The computational effort for the simulation of 20  wave groups (10 minutes of physical time) in parallel computation on 8 CPUs was approximately one month.

\section{Experimental and model results for the outer flow hydrodynamics}
\label{sec:Results}
In this section results for surface elevations, outer flow (positions well clear of the boundary layer) velocities and turbulence from both the experiments and the numerical model will be presented and compared. 

\subsection{Surface elevations}
\label{sec:surface}
\begin{figure*}
	\centering
    \includegraphics[trim=1cm 1.5cm 0cm 1.2cm, clip=true, scale=1]{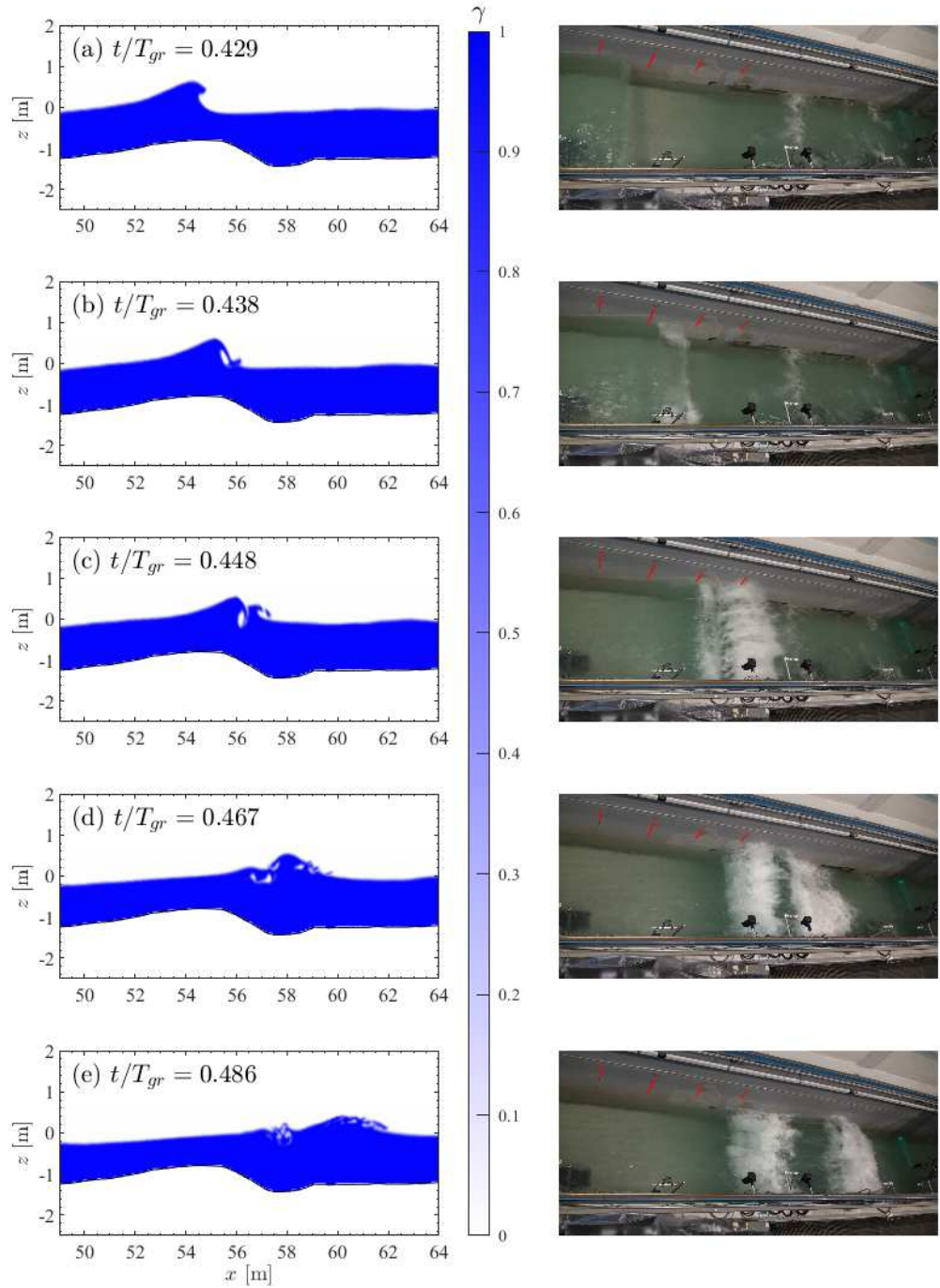}
\caption{Breaking sequence of the model (left) and the experiments (right). The top left hand side of the video frames correspond to $x\approx 53.5$ m and at the right hand side the laser from the LDA can be seen at $x=64$ m. Red-lines have been added on the video frames to highlight the one m spaced lines drawn on the wall.  The leftmost red line corresponds to $x=54.4$ m. Note the difference in scaling between the vertical and horizontal axis for the model results.}    
    \label{fig:BreakingSnapShots}
 \end{figure*}

Figure \ref{fig:BreakingSnapShots} shows snapshots of the breaking sequence of the third wave in the group from the model (left) and video frames from the experiment (right). The top left hand side of the video frames corresponds to $x \approx 53.5$ m, whereas the modelled results start at $x=49$ m. In this figure and for the remainder of this article, $t/T_{gr}=0$ corresponds to the arrival of the group at $x=49$ m, to be consistent with \cite{vanderZandenetal2019}.   In Figure \ref{fig:BreakingSnapShots}a the wave has just started to overturn and in Figure \ref{fig:BreakingSnapShots}b the plunging jet can be seen just touching the water. In Figure \ref{fig:BreakingSnapShots}c the plunging jet entrains a lot of air and pushes up a wedge of water in front. Figure \ref{fig:BreakingSnapShots}d shows the generated splash up and Figure \ref{fig:BreakingSnapShots}e shows the breaking bore that propagates towards the shore, while the water near the surface above the bar trough contains a mixture of air and water. 
The modelled and experimental breaking sequences shown here are qualitatively very similar. The bichromatic group consisted of seven short waves, the five largest of which broke above the bar (each as a plunging breaker). The breaking sequences shown in Figure \ref{fig:BreakingSnapShots} are typical for all of the five plunging waves in the  group, although the cross-shore position of the break point and plunge point varied for the individual waves in the group.

Table 1 shows the measured and modelled horizontal positions of the break points and plunge points as well as the breaker type of each wave in the group. Following the classification from \cite{SmithKraus1991}, the break point has been defined as the horizontal position where the wave starts to overturn and the plunge point as the horizontal position where the jet first hits the water. The break and plunge point of the experiments have been determined from video recording similar to the images shown in Figure \ref{fig:BreakingSnapShots}, by using the red vertical lines on the wall, which are positioned with 1 m intervals. The exact position of the break and plunge points in the experiments are naturally hard to evaluate due to the placement of the camera and the limited grid on the wall. Therefore the break and plunge points have been given as a range rather than a fixed position.
 \begin{table}
\centering
\caption{Comparison between the experimental and modelled  horizontal position of the break points, plunge points and breaker type} 
\begin{footnotesize}
\label{Tab:resultat}
\begin{tabular}{ccccccc} \hline
Wave no. & Break Point [m] & Break Point [m]&  Plunge Point [m] &  Plunge Point [m]&Breaker type & Breaker type\\ 
& Experiments & Model & Experiments &Model & Experiments & Model\\
\hline

1 & non breaking&  non breaking&  non breaking&  non breaking &  non breaking& non breaking\\ 
2 & 55.5--56.4 & 56  & 56.5--57 & 57.5 &plunging&plunging \\
3 & 53.5--54.5 & 54.3 & 55.5--56.5 & 55.8 &plunging&plunging\\
4 & 53.5--54.5 & 53 & 56--56.5 & 54.8 &plunging&plunging \\
5 & 54.4--55 & 54.2  & 56--56.5 & 55.8 &plunging&plunging \\
6 & 55--55.5 & 58--60 & 56.4--57 & -  &plunging&spilling\\
7 & non breaking & non breaking  &  non breaking& non breaking & non breaking& non breaking \\
\hline
\end{tabular}
\end{footnotesize}
\end{table}
The smallest waves (waves 2 and 6) broke furthest onshore while the larger waves broke further offshore. 
The model, in general,  captures both the break point and the plunge point of waves 2-5 well, although the modelled fourth and fifth waves broke slightly further offshore compared to the experiments.
In the experiments the sixth wave also broke over the bar as a plunging breaker. In the model this wave appeared to be close to overturning over the bar crest ($x=55$ m), but then lost some of its steepness on the lee side of the bar before breaking as a spilling breaker while it propagated out of the bar trough which explains the big difference in break point between the model and the experiments for this wave (Table \ref{Tab:resultat}). 

\begin{figure}[h!]
	\centering
    \includegraphics[trim=0cm 0.6cm 0cm 1cm, clip=true, scale=1]{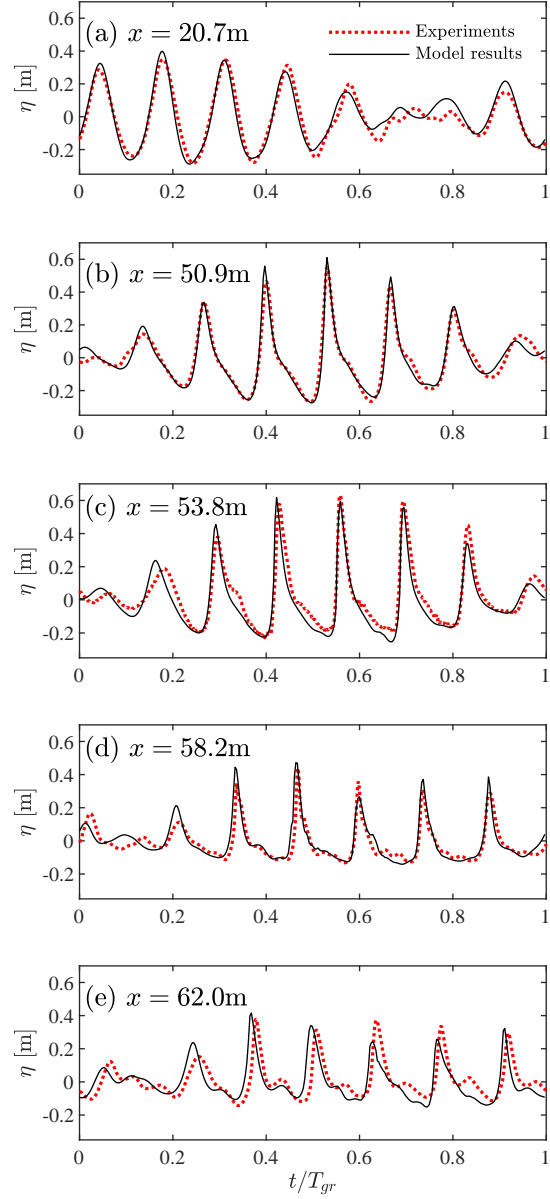}
\caption{Time series of the phase-averaged water surface elevations at five different cross shore positions}    
    \label{fig:SurfaceElevationTimeSeries}
 \end{figure}
 
Figure \ref{fig:SurfaceElevationTimeSeries} shows the measured and modelled phase-averaged surface elevations at five locations: in the flat part of the flume (Figure \ref{fig:SurfaceElevationTimeSeries}a), at a shoaling position (Figure \ref{fig:SurfaceElevationTimeSeries}b), at the bar crest (Figure \ref{fig:SurfaceElevationTimeSeries}c), at the bar trough which can be considered the outer surf-zone (Figure \ref{fig:SurfaceElevationTimeSeries}d) and in the inner surf zone (Figure \ref{fig:SurfaceElevationTimeSeries}e). From the flat part of the flume towards the shoaling position (Figure \ref{fig:SurfaceElevationTimeSeries}a-b) the wave skewness and asymmetry have increased with crests being much higher than the troughs and the waves having a characteristic saw-tooth shape, and this shape continues through the rest of the flume.
The model and experimental results are similar for all five cross-shore positions, but some notable differences can be observed. In the incoming wave signal (Figure \ref{fig:SurfaceElevationTimeSeries}a) large differences can be seen between the smallest waves in the  group, and this difference persists at the other locations along the flume (Figure \ref{fig:SurfaceElevationTimeSeries}b-e). Although the measured and modelled surface elevations are reasonably in phase near the bar trough (Figure \ref{fig:SurfaceElevationTimeSeries}d), the modelled surface elevations lead the experimental observations in the inner surf zone (Figure  \ref{fig:SurfaceElevationTimeSeries}e). This implies a slight overestimation in wave propagation speed in the model. The sixth wave in the  group at the bar crest (Figure \ref{fig:SurfaceElevationTimeSeries}c, $t/T_{gr} = 0.82$) is higher in the experiment than in the model, explaining why this particular wave did not plunge over the bar in the model in contrast to the experiments. This difference may be explained by a more pronounced bound long wave in the model relative to the experiments (this can most clearly be seen in the forthcoming Figure \ref{fig:FreestreamVel}), which is probably related to the the difference in wave generation between model and experiments (the experiments used first order generation  using a wedge without active absorption, whereas the model used second order generation via a relaxation zone, thereby giving active absorption). Attempts were made to generate the waves differently in the model, e.g. first order generation rather than second (though maintaining active absorption for stability) and changing the phase of the two frequency components. None of these attempts caused all five waves to break as plungers over the bar.

We will now focus on the main statistics of the wave surface elevations by looking specifically at maximum and minimum phase-averaged surface elevations ($\eta_{max}$ and $\eta_{min}$), the root mean square (r.m.s.) of the phase-averaged surface elevations ($\eta_{rms}$) and the wave skewness, which is a measure for the ratio between crest and trough heights, as

\begin{equation}
S\!k(\phi)=\frac{\langle\phi^3\rangle}{\phi_{rms}^3}
\end{equation}
where $\phi$ can represent any quantity e.g surface elevations $\eta$ or velocity $u$ and the angular brackets represents averaging over the entire group according to

\begin{equation}
\langle \phi \rangle = \frac{1}{T_{gr}} \int_0^{T_{gr}} \phi dt
\end{equation}
Finally, we will also look at the wave asymmetry, which is a measure for the ratio between water surface steepness at the wave front versus the wave rear \citep{ElgarGuza1985}

\begin{equation}
A\!s(\phi)=\frac{\langle\mathcal{H(\phi)}^3\rangle}{\phi_{rms}^3}.
\label{eq:asym}
\end{equation}
Here, $\mathcal{H}$ is the Hilbert transform. For sawtooth-shaped waves with a steep front and mild rear slope, $As$ is negative. The above formulations are used to calculate the skewness and asymmetry over the phase averaged wave group.

\begin{figure*}
	\centering
    \includegraphics[trim=1cm 1.5cm 0cm 1.1cm, clip=true, scale=1]{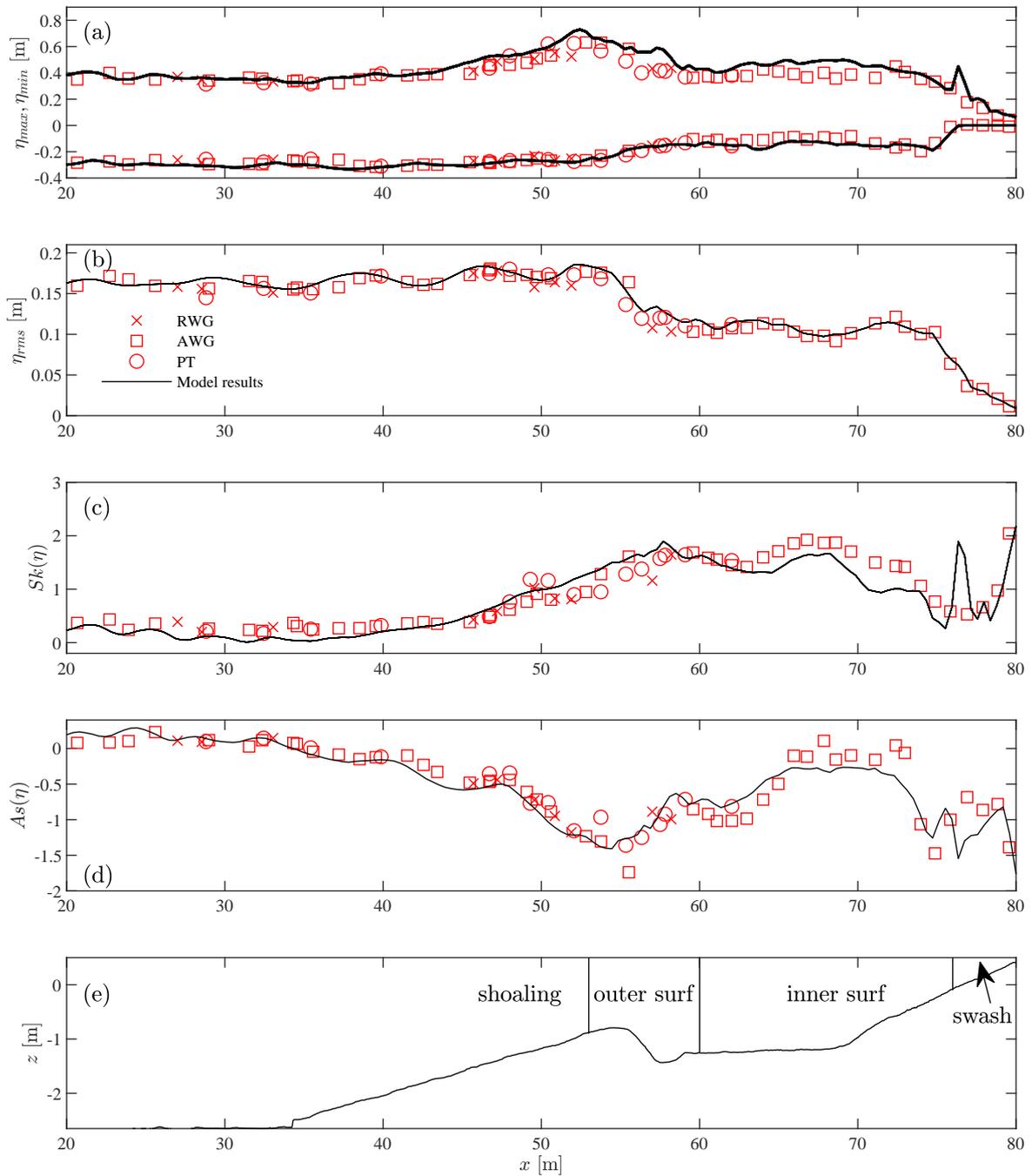}
\caption{Comparison between experimental and model results for (a) surface elevation envelope, (b) rms of the surface elevations, (c) wave skewness, (d) wave asymmetry and (e) the bed profile with indication of the different cross-shore zones. }    
    \label{fig:WaveEnvelope}
 \end{figure*}

Figure \ref{fig:WaveEnvelope}a shows the modelled maximum and minimum phase-averaged surface elevations along the length of the flume compared against the experiments. The model captures both the water surface maxima and minima in the flat part of the flume ($x<34$ m) and during shoaling ($34$ m $<x< 53$ m), and is able to capture the point where the wave heights start to decrease due to breaking ($x\approx 53$ m) marking the start of the outer surf zone. In the inner surf zone, where the waves have turned into breaking bores ($60$ m $ < x<76$ m) crest heights are slightly overestimated, compared to the AWG measurements, which might be due to the applied de-spiking routine that slightly smooths the measured wave crests \citep{vanderZandenetal2019}. At $x \approx 74$ m, the second decay in wave height is also captured. The largest difference between the modelled results and the experiments can be seen in the swash zone ($x>76$ m, the region alternately inundated and exposed by
flow uprush and backwash) where the modelled results show an additional small peak ($x\approx 76$ m). This peak results from the splash up of one of the waves in the  group, which breaks for a second time in the swash zone. In the experiments the waves also shoaled and broke again in the swash zone, but a splash up as predicted by the model was not observed.

Figure \ref{fig:WaveEnvelope}b illustrates that $\eta_{rms}$ is almost constant in the flat part of the flume and during shoaling in both the model and the experiments. This indicates, similar to the suggestion in \cite{vanderZandenetal2019}, that the increase in wave height shown in Figure \ref{fig:WaveEnvelope}a relates to an increase in the wave skewness. In the flat part of the flume and during shoaling an oscillatory pattern can be detected in the experimental $\eta_{rms}$, which is reasonably captured by the model, although the wave length of the oscillations in the simulations is slightly longer than in the experiments. A possible explanation for this discrepancy is the overestimation in propagation speed by the model in the inner surf zone probably resulting in different reflection from the beach.

Both in the experiments and in the modelled results the skewness increases during shoaling and remains at a relatively constant high level in the surf zone, before decaying from $x>68$ m (Figure \ref{fig:WaveEnvelope}c). In the model a sudden increase in skewness can be seen at $x\approx 76$ m, which again corresponds to the splash up of one of the waves. 
The magnitude of the asymmetry increases during shoaling (Figure \ref{fig:WaveEnvelope}d), but drops significantly between the points where the wave starts to overturn ($x\approx 53$ m) and the plunging jet hits the water ($x \approx 55$ m). Near the swash zone, with its second shoaling and breaking sequence, the magnitude of the asymmetry increases again. Overall, the cross-shore distribution of the skewness and asymmetry is well captured by the model.

In this section no direct comparisons of surface elevations were made with results from the standard (non-stabilized) model, but for completeness we would like to state that the surface elevations with the standard model were almost identical to the stabilized model. The over-production of turbulence, to be shown in the coming subsections, did therefore not have any significant influence on the surface elevations.

\subsection{Outer flow velocity}
\label{sec:outer}

\begin{figure*}[ht]
	\centering
	
    \includegraphics[trim=1cm 0cm 0cm 0cm, clip=true, scale=0.9]{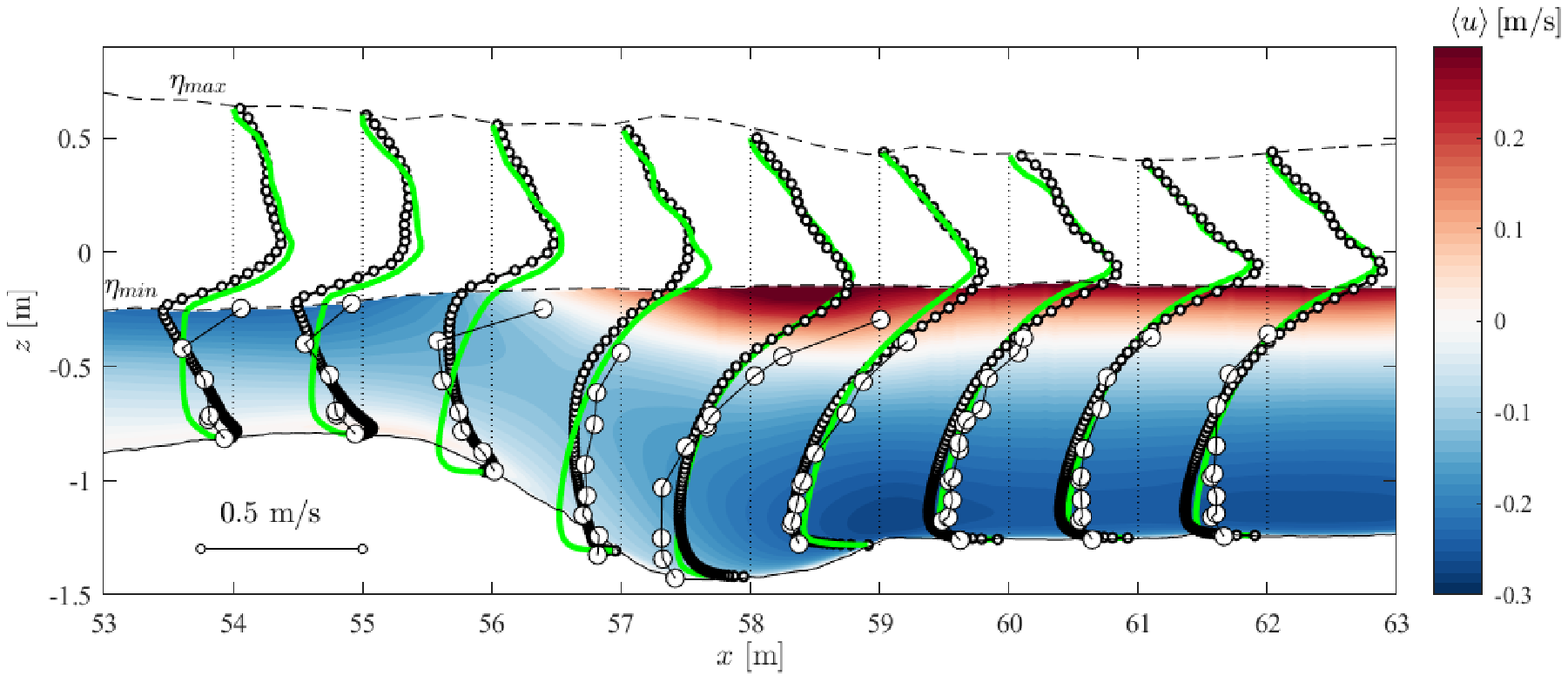}
\caption{Spatial distribution of the time-averaged cross-shore velocity of the experiments (large circles), from the model (small circles and colored velocities) and using a standard (non-stabilized, but including buoyancy modification) model (green lines).}    
    \label{fig:UndertowModel}
 \end{figure*}

Figure \ref{fig:UndertowModel} shows a comparison of the spatial distribution of the time-averaged cross-shore velocity (the undertow) between the model (small circles and colors) and the experiments (large circles). 
For reference, the figure also includes model results using a standard  (non-stabilized, still including the buoyancy production term) two-equation turbulence model (i.e. setting $\lambda_2=0$, here shown with green lines). 
The experiments show two distinctly different undertow profile shapes in the shoaling region and in the surf zone. In the shoaling region and the very outer surf zone ($x < 56$ m) the undertow is strongest far away from the bed, whereas in the bar trough and inner surf zone ($x > 58$ m) the strength of the undertow is strongest near the bed. This finding is similar to regular waves propagating and breaking over a bar \citep{vanderAetal2017}, and also similar to what was seen in the small-scale experiments of \cite{TingKirby1994} for a plane sloping beach. The breaker bar naturally plays an important role for the mean velocity profiles, but since this particular undertow profile evolution shows up with and without a bar, it seems that the cross-shore distance relative to the breaking point is most important for the shape of the undertow profile. The qualitative difference in the undertow profile as well as the transition in profile shape from one region to the other is well captured by the model. On the other hand, the standard turbulence model does not capture the difference in profile shape nor the transition in shape between $x = 56$ -- $58$ m and instead, predicts a similar profile shape from the shoaling zone to the inner surf zone. This erroneous behaviour can be considered typical of standard RANS models in the shoaling and outer surf zone, as demonstrated in \cite{Brown2016} who tested several RANS turbulence models for breaking waves. In \cite{LarsenFuhrman2018} this behaviour was attributed to the over-production of turbulence in the pre-breaking region, which un-physically increases the flow resistance in the upper part of the flow and forces the undertow to maintain the same shape as in the surf zone. 

The strength of the undertow in the inner surf-zone ($x \geq 60$ m) is overestimated by the model compared to the experiment (the maximum magnitude of the offshore directed flow is overestimated with a factor 1.2-1.5 in this region). This is similar to the simulations of the \cite{TingKirby1994} experiments with the same model presented in \cite{LarsenFuhrman2018} and has been widely observed for RANS simulations, see e.g. \cite{Brown2016}.
 
 \begin{figure}[ht!]
	\centering
    \includegraphics[trim=0cm 0.9cm 0cm 1.2cm, clip=true, scale=1]{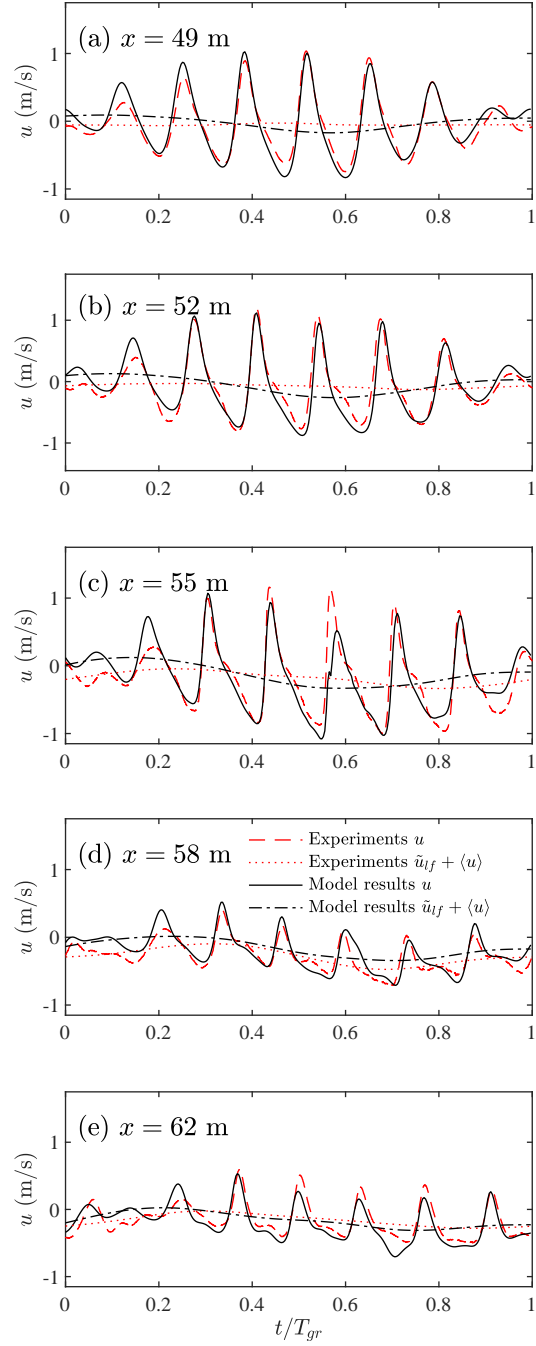}
\caption{Time series of horizontal phase-averaged velocities at $\zeta=z-z_{bed}\approx 0.4$ m at five different cross-shore positions for the model and the experiments.}    
    \label{fig:FreestreamVel}
 \end{figure}
 
Having shown a reasonable match between the modelled and experimental time-averaged velocity profiles, we will now look at the temporal development of the free-stream velocity signal at different cross-shore positions.  Figure \ref{fig:FreestreamVel}  shows time series of the phase-averaged free-stream velocities at an elevation  $\zeta=z-z_{bed}\approx 0.4$ m at five different cross-shore positions, corresponding to two shoaling positions (Figure \ref{fig:FreestreamVel}a,b), the bar crest (Figure \ref{fig:FreestreamVel}c), the bar trough (Figure \ref{fig:FreestreamVel}d) and a position in the inner surf zone (Figure \ref{fig:FreestreamVel}e). Included in the figure is also the sum of the low frequency and group-averaged velocity which comes from a decomposition of $u$ into group-averaged ($\langle u\rangle$), high frequency oscillating ($\tilde{u}_{hf}$) and low frequency oscillating ($ \tilde{u}_{lf}$) components

\begin{equation}
u = \tilde{u}_{hf}  + \tilde{u}_{lf} + \langle u\rangle
\end{equation}
where a cut-off frequency of 0.1 Hz (approximately half the frequency of the short waves) has been used to separate low frequency and high frequency components.
Velocities in the shoaling region (Figure \ref{fig:FreestreamVel}a,b) are skewed and show a reasonable match between the experiments and the model. Similar to the surface elevations, the smallest waves in the  group are not captured well by the model, and especially the maximum value of the positive velocities of the first two waves in the  group are larger in the simulations. Furthermore, the model shows slightly larger magnitudes of the negative velocities beneath the troughs  of the three waves in the center of the  group. These differences can largely be explained by the model having a clear low frequency contribution to the velocity in these positions, in contrast to the experiments.

At the bar crest (Figure \ref{fig:FreestreamVel}c) the velocities follow a characteristic sawtooth shape. This is as expected as this cross-shore position corresponds to initiation of breaking for most of the waves. The experiments and modelled results generally compare well in this position. The difference between model and experiments for the smallest waves in the  group are more pronounced at this position than in the shoaling zone, with the modelled first wave having a larger velocity amplitude and the modelled seventh wave having a lower velocity amplitude compared to the experiment. Furthermore, both the maximum and minimum velocity of the center wave ($t/T_{gr} \approx 0.52$--$0.58$) are lower in the model compared to the experiment. This relates to the wave breaking slightly further offshore in the model compared to the experiment (Table \ref{Tab:resultat}). 

In the bar trough (Figure \ref{fig:FreestreamVel}d) a gradual increase in offshore directed velocity with each passing broken wave can be seen, creating a clear low-frequency velocity fluctuation at the wave  group time scale in both the experiments and the model. In the inner surf zone (Figure \ref{fig:FreestreamVel}e),  both the high frequency and the low frequency contributions to the free stream velocity signal is reasonably well captured by the model, although the overestimation of the undertow strength by the model is visible.

\subsection{Outer flow turbulence}
\label{sec:outerk}
In this section the modelled TKE will be compared to the experiments. In \cite{Jacobsenetal2014} comparison between experimental and modelled TKE profiles was done taking into account both the modelled turbulence as well contributions from wave to wave variation, which could be interpreted as part the part of the turbulence being resolved by the model (e.g. the roller). In this paper only the modelled TKE will compared, as it has been confirmed that the modelled TKE is completely dominating in regions where the model is compared to the experiments. 

\begin{figure*}[ht!]
	\centering
    \includegraphics[trim=0cm 0.cm 0cm 0cm, clip=true, scale=0.9]{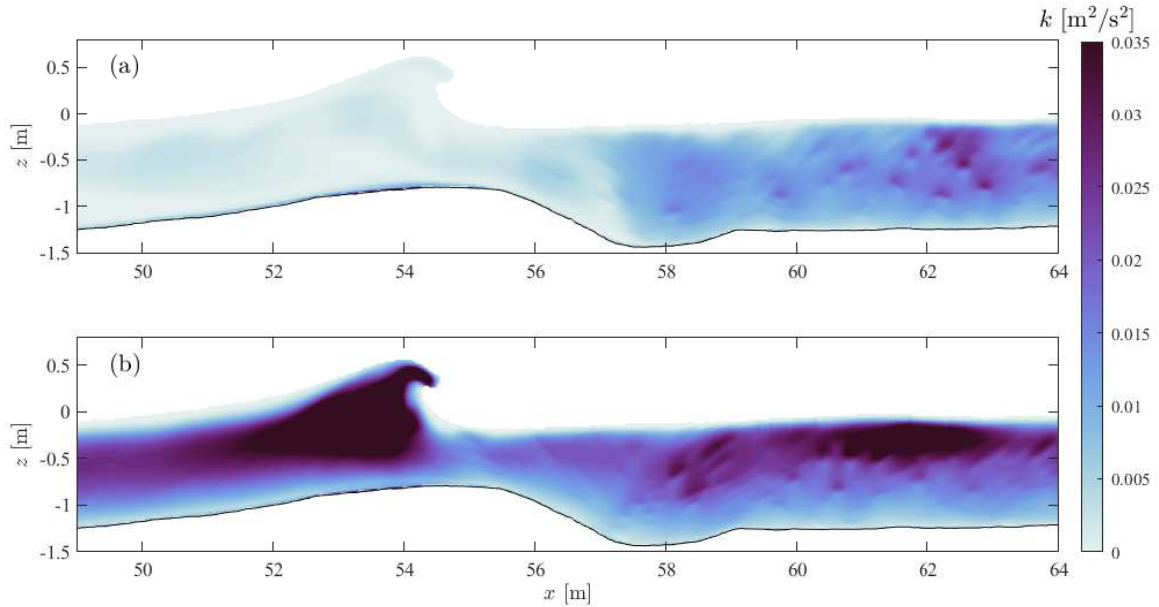}
\caption{Instantaneous snapshot of the spatial distribution of TKE upon wave breaking using a (a) stabilized and (b) standard two-equation turbulence model. In both cases buoyancy modification is included.}    
    \label{fig:InstabilityExample}
 \end{figure*}

\begin{figure*}[ht!]
	\centering
    \includegraphics[trim=1cm 0.cm 0cm 0cm, clip=true, scale=0.9]{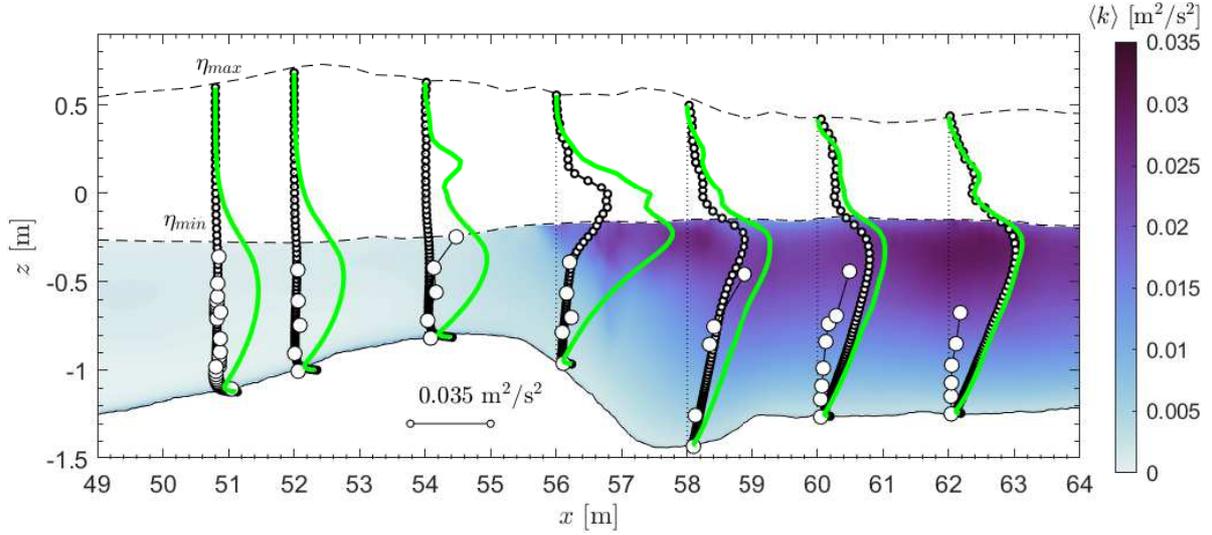}
\caption{Spatial distribution of the time-averaged TKE of the experiments (large circles), from the model (small circles and color contours) and using a standard (non-stabilized) model (green lines).}    
    \label{fig:MeanK}
 \end{figure*}

Figure \ref{fig:InstabilityExample} shows a snapshot of the TKE field of the third wave in the  group, modelled using both a stabilized turbulence model (Figure \ref{fig:InstabilityExample}a) and a standard model (Figure \ref{fig:InstabilityExample}b). The wave is on the verge of breaking to clearly illustrate the shoaling and surf zone regions. 
The stabilized model results in high levels of TKE only in the surf-zone  and wave boundary layer as would be expected.  The standard (non-stabilized) model on the other hand, produces very high TKE levels prior to breaking (even higher levels than in the surf zone), which is in sharp contrast to many experimental observations that have shown low levels of turbulence prior to breaking \citep[see e.g.][]{TingKirby1994, vanderAetal2017,vanderZandenetal2019}. It is emphasized that the above standard (non-stabilized) model (Figure \ref{fig:InstabilityExample}b, as well as the green lines in Figure \ref{fig:UndertowModel} and the forthcoming Figure \ref{fig:MeanK}) includes buoyancy modification similar to that proposed in \cite{Devolderetal2017}. This further demonstrate that, while this modification creates a local sink of turbulence at the air-water interface, it does not solve the over-production of turbulence elsewhere. This is consistent with results shown in \cite{LarsenFuhrman2018}, their Figures 3, 6a,b and 12, which demonstrate that (traditional, non-stabilized) models with this term active still lead to exponential growth and pronounced over-production of TKE and eddy viscosity throughout the nearly potential flow region. It is the modification to the eddy viscosity in equation \eqref{eq:wtildeNew}, combined with the modified production term for $\omega$ in equation \eqref{eq:Pomega}, as proposed and analyzed by \cite{LarsenFuhrman2018}, which formally stabilizes the model and prevents the substantial over-production of turbulence prior to breaking. 

Figure \ref{fig:MeanK}a presents the time-averaged TKE of both the model and the experiments. This figure also includes the results using the standard two equation turbulence closure.  
Using a formally stabilized turbulence model, the TKE levels at shoaling and wave breaking locations ($x=50.7-56$ m) are generally low and correspond well with the measured TKE. In contrast, using a standard two-equation turbulence model yields TKE levels in these regions that are of similar magnitude as in the surf-zone  and that are several orders of magnitude larger than the measured levels. 
In the inner surf zone both the stabilized and standard (non-stabilized) model overestimates the TKE. The stabilized model overestimates the TKE with approximately a factor 3-5 in the inner surf zone. The slightly larger overestimation for the non-stabilized model can be explained by the waves arriving at the surf zone with severely overestimated turbulence levels advecting additional turbulence into the surf zone. By design, as it is only intended to prevent the non-physical growth in the potential flow (pre-breaking) region, the stabilized model has little effect in the inner surf zone, where rotation rates and strain rates are of the same order of magnitude (the new limiter is thereby effectively turned off). 

\begin{figure*}[ht!]
	\centering
    \includegraphics[trim=1cm 1cm 0cm 0.2cm, clip=true, scale=1]{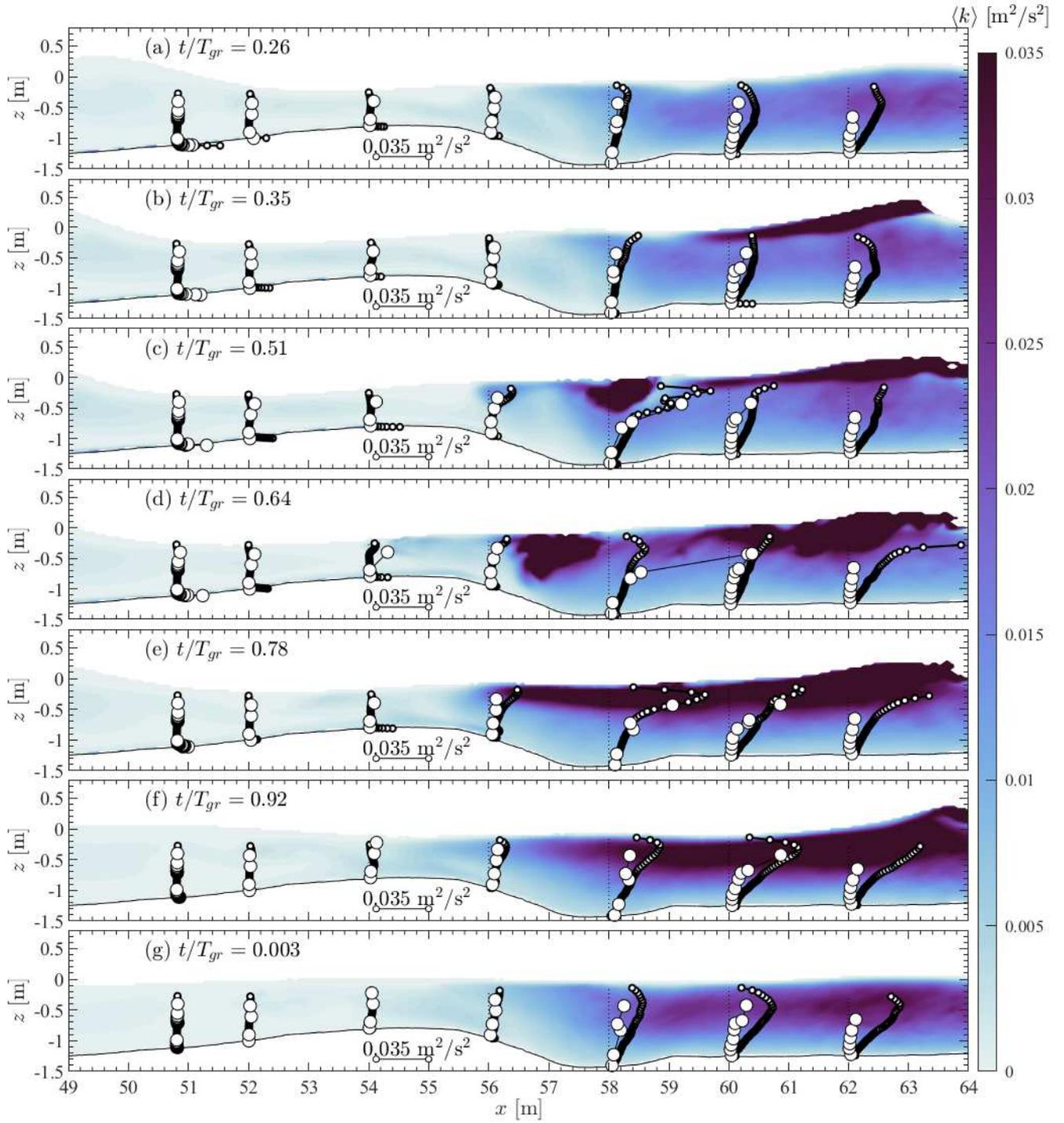}
\caption{Comparison of experimental (large circles) and modelled (small circles and color contours) phase-averaged TKE during the passing of the seven waves in the  group.}    
    \label{fig:phaseK}
 \end{figure*}

To further analyze the model performance and the modelled spatiotemporal variations in TKE, Figure \ref{fig:phaseK} shows the phase-averaged TKE of the model and the experiments at seven selected times, corresponding to the passing of each of the seven individual wave crests of the wave group at the end of the measurement area ($x=64$ m).  For an analysis of the physical mechanisms (i.e., advective and diffusive transport and local production and dissipation of turbulence) that drive the spatial and temporal variations of TKE, we refer to \cite{vanderZandenetal2019}. The present section focuses primarily on the behavior of turbulence in the model as well as on the agreement between model and data.   
In Figure \ref{fig:phaseK}a the TKE levels are generally at their lowest, since the two most recent waves did not break in the test section. Some turbulence, originating from wave breaking more than two wave cycles before, still persists in the bar trough and inner surf zone ($x>58$ m). This is consistent with the measurements and implies that the wave breaking turbulence requires multiple wave cycles to fully decay. In general, a good match between the modelled and the experimental TKE can be seen at this phase, although the model overestimates the turbulence levels in the innermost position ($x=62$ m). 

In Figure \ref{fig:phaseK}b the first breaking wave of the group has passed the bar and arrives to the inner surf zone at $x = 63$ m as a surf bore. The crest of the bore contains high levels of TKE. Behind the bore, high levels of TKE are observed in the upper part of the flow, but further down the water column the TKE levels have not increased significantly compared to Figure \ref{fig:phaseK}a. This behaviour is consistent in model and measurements. 

In Figure \ref{fig:phaseK}c another breaking bore can be seen at the right side of the figure. Similar to the previously shown phase, the breaking bore contains high levels of TKE and is leaving a trace of highly turbulent flow in the upper water column, which in this case invades further down into the water column compared to the previously shown phase. In contrast to the previous phase, high levels of TKE are now also present in the bar trough in both the experiments and in the model. The high TKE levels in the trough can firstly be explained by the wave plunging further offshore and secondly by the plunger being stronger, thereby penetrating further into the water column. For this phase some TKE has been advected offshore by the undertow resulting in higher turbulence at $x=56$ m, a process that is also captured by the model. 

In Figure \ref{fig:phaseK}d the largest wave in the  group has passed. The image is very much the same as in the previous phase. The turbulence from the breaking bore is extending further down the water column, in the bar trough a highly turbulent zone remains and the undertow is convecting this turbulence over the bar crest. The TKE levels at the uppermost position in the bar trough are significantly higher in the experiments than in the model. This can be explained by the  breaking location in the model being slightly more offshore compared to the experiments (Table \ref{Tab:resultat}). An area with high turbulence levels is present in the model just offshore of $x=58$ m (Figure \ref{fig:phaseK}d), which support this explanation. 

In  Figure \ref{fig:phaseK}e-f, with the passing of the fourth and fifth breaking wave, TKE is extending even further down the water column and a good match between the experiment and the model can be seen. Turbulence becomes increasingly more uniformly distributed during this stage of the wave  group cycle. Finally, in Figure \ref{fig:phaseK}g the seventh wave (non-breaking) has just passed. The turbulence has reduced significantly compared to the previous phase. The TKE in the experiments decays somewhat faster than in the model, especially at $x=56$--$62$ m.

\section{Boundary layer characteristics}
\label{sec:NearBed}
\subsection{Boundary layer flow}
\label{sec:BL}
In the experiments detailed measurements of the boundary layer were performed using an LDA at $x=50.77$ m.  The measurements were obtained at 15 elevations logarithmically spaced between $\zeta = z - z_{bed}$ = 0.001 and 0.125 m, where $z_{bed}$ was defined in the measurements as the top of the roughness elements. Additionally, near-bed velocities were measured with the LDA at one elevation ($\zeta \approx 0.025 $m) at several cross-shore positions. In this section we will focus on the detailed measurements in the shoaling position, but the results from the other positions will be discussed in Section \ref{sec:BLk}.
The near-bed LDA data were processed in the same way as the outer flow data, following procedures described in \citet{vanderZandenetal2019}.

As shown previously (Figure \ref{fig:FreestreamVel}), the free-stream velocity signals of the model and experiments are similar, but differ slightly in terms of magnitude, phase and amplitude of the long wave. Therefore, to make the most fair comparison and not carry already established differences into this comparison, the model and experimental profiles will be compared for $ \tilde{u}_{hf} $ rather than $ u$. 

\begin{figure*}[ht!]
	\centering
    \includegraphics[trim=0.7cm 1.5cm 0cm 1.3cm, clip=true, scale=0.95]{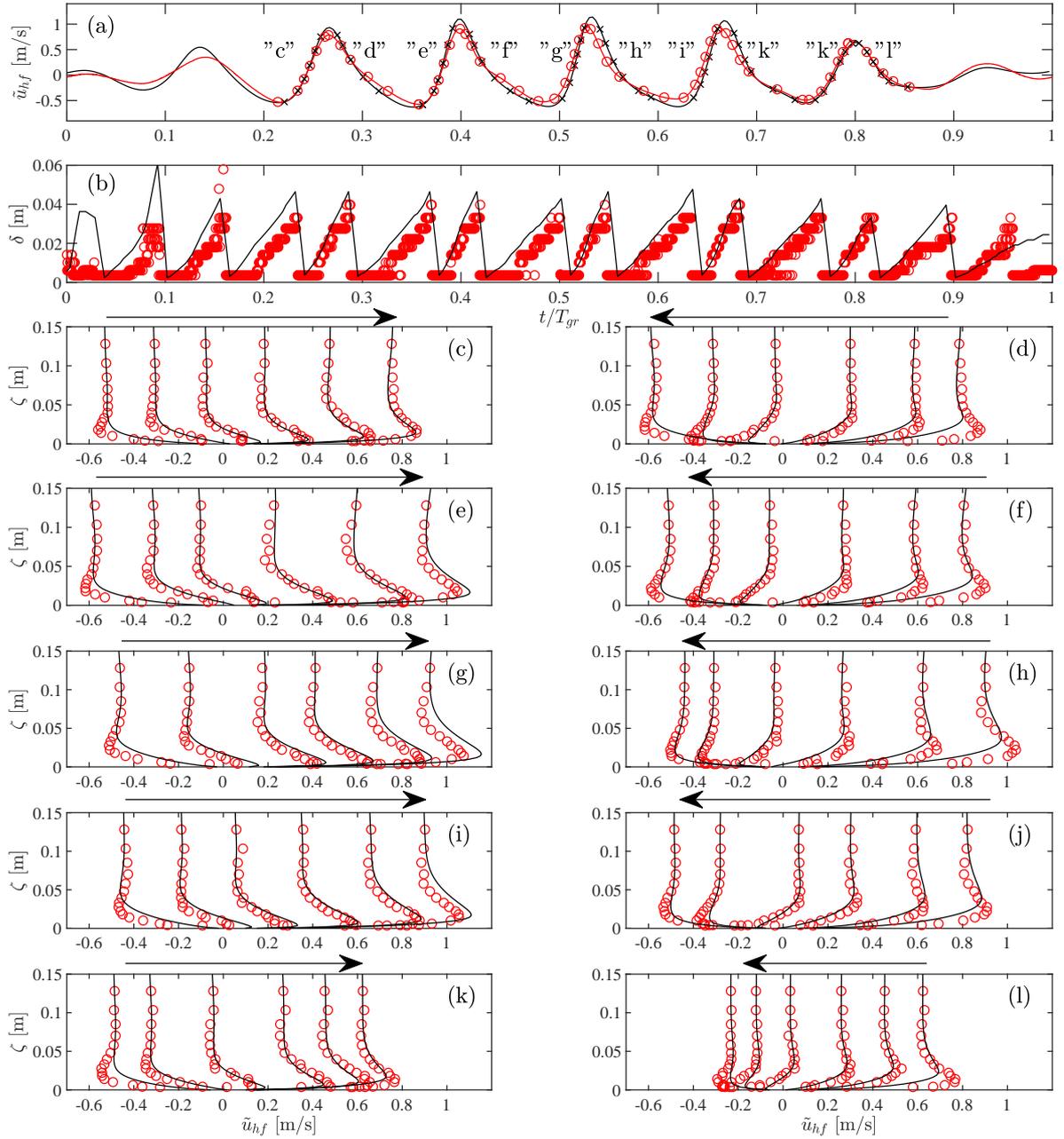}
\caption{(a) Phase-averaged high frequency component velocity signal at $x=50.77$ m and $z-z_{bed} = 0.125$ m; (b) boundary layer thickness; (c-l) Vertical profiles of the phase-averaged modelled (solid) and measured (circles) high frequency component velocities at selected phases. The selected phases are referenced in (a) by cross (model) and circle (measurements) markers. The arrows mark the direction of the acceleration.}    
    \label{fig:DetailedBL}
 \end{figure*}

Figure \ref{fig:DetailedBL} shows time series of $ \tilde{u}_{hf} $ at $x=50.77$ m  and $\zeta = 0.125$ m (Figure \ref{fig:DetailedBL}a) as well as the development of the boundary layer thickness of both the experiment and the model. Here the boundary layer thickness has been determined following the approach by \cite{LarsenFuhrman2019b} as the first vertical position where the following is exceeded

\begin{equation}
\frac{\zeta}{|u_{hf}|}\frac{\partial |u_{hf}|}{\partial \zeta}<0.03
\end{equation}
This formulation has been chosen over more typical formulations like the position where the velocity exceeds 95 $\%$ of the free stream velocity or the position of the velocity overshoot (i.e. where the velocity gradient changes sign). The first of these approaches (with a fixed free-stream height) would lead to an underestimation of the boundary layer thickness in situations with a large velocity overshoot, and the second approach would fail in situations where no overshoot is present and the velocity gradient does not change sign.

Figure \ref{fig:DetailedBL} likewise includes the development of the near-bed profile of $ \tilde{u}_{hf} $ for the five largest waves in the wave  group (Figure \ref{fig:DetailedBL}c-l), from both the model and the experiments. The modelled and measured velocities profiles shown are synchronized at $\zeta = 0.125$ m. The phases and free-stream velocities chosen for comparison are shown in Figure \ref{fig:DetailedBL}a.

One of the most striking features of the velocity profiles shown in Figure \ref{fig:DetailedBL} is the very large velocity overshoot during the acceleration part of the onshore phases (part of wave cycle where the velocity is onshore directed), shown in the right-most profiles in Figure \ref{fig:DetailedBL}c,e,g,i,k.
During the acceleration part of the offshore phases (part of wave cycle where the velocity is offshore directed), shown in the left-most profiles in Figure \ref{fig:DetailedBL}d,f,h,j,l, the flow acceleration is of much smaller magnitude, resulting in a smaller overshoot compared to the crest half-cycle. Additionally, converging/diverging effects may contribute, creating large near-bed velocities during the onshore phases compared to the offshore phases, as shown in \cite{Sumeretal1993} and \cite{Fuhrmanetal2009a}.
The model generally captures the shape of the velocity profiles in the wave boundary layer with reasonable accuracy. For the acceleration part of the onshore phases, a good match between the measured and modelled overshoot magnitude and elevation is achieved for the second wave (Figure \ref{fig:DetailedBL}c). For waves three, four and five the elevation of the overshoots are well captured but the magnitudes of the overshoots are slightly overestimated (Figure \ref{fig:DetailedBL}e,g,i). This is attributed to an overestimation of the orbital excursion and to the mean fluid acceleration between trough and crest phase by the model for these three waves (Figure \ref{fig:DetailedBL}a). For waves four and five (Figure \ref{fig:DetailedBL}g,i), the velocities above the maximum velocity overshoot are generally higher in the model compared to the experiments. This could be due to an overestimation of the turbulence levels in the model in the upper part of the boundary layer (see Section \ref{sec:BLk}), leading to an increased vertical exchange of momentum. Finally, for wave six the velocity overshoot is slightly underestimated (Figure \ref{fig:DetailedBL}k).
During the deceleration part of the onshore phases (right-most profiles in Figure \ref{fig:DetailedBL}d,f,h,j,l) the modelled boundary layer has grown slightly faster in the model compared to the experiments, which can also be seen in Figure \ref{fig:DetailedBL}b). This is attributed to an overestimation of the orbital excursion in the model which can be expected to generate large boundary layer thickness \citep{FredsoeDeigaard1992}.
 For the acceleration part of the offshore phases (left-most profiles of Figure \ref{fig:DetailedBL}d,f,h,j,l) the velocity overshoot tends to be underestimated in terms of magnitude and slightly overestimated in terms of height by the model. This is consistent with the image in Figure \ref{fig:DetailedBL}b where it can be seen that the model has a tendency to slightly overestimate the boundary layer thickness, and that this is most pronounced during the deceleration part of the offshore directed phases. This can again be attributed to overestimated orbital excursion, but additionally it could indicate that the chosen $k_s$ is too high.

\begin{figure}[ht!]
	\centering
    \includegraphics[trim=0cm 0cm 0cm 0cm, clip=true, scale=1]{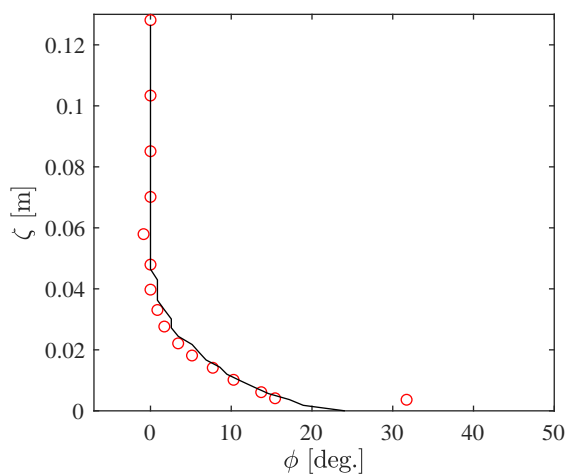}
\caption{Vertical distribution of the phase lead at $x=50.77$ m for both the model (line) and the experiment (circles).}
    \label{fig:BL5077phaseDiff}
 \end{figure}

Consistent with previous research \citep{Sleath1987,Jensenetal1989,vanderAetal2011} a clear phase lead between the near bed flow and the free-stream velocity can be seen. This is for instance evident in Figure \ref{fig:DetailedBL}e where the near bed velocity (in the third profile from the left) is positive ($ \tilde{u}_{hf} \approx 0.2$ m/s) whereas the free-stream velocity is still negative ($ \tilde{u}_{hf} \approx -0.1$ m/s). To quantify this phase lead, the velocities at each vertical position were cross correlated with the free-stream velocity signal and were normalized using the mean period $T_m$ for phase reference. 
Figure \ref{fig:BL5077phaseDiff} shows the vertical distribution of phase leads of both the experiment and the model. 
The phase lead starts at $\zeta \approx 0.035$ m and increases progressively downwards in both the experiments and in the model, up to a maximum near the bottom. In the experiments the near-bed phase lead of 32$^\circ$ agrees well with Figure 6 from \cite{vanderAetal2011}, where phase leads from various oscillatory flow experiments are shown as a function of $a/k_s$. In the model, the near-bed phase lead is not as large as in the experiments, but overall the vertical trend is highly similar. The discrepancy nearest to the bottom could also be a result of heterogeneous bed roughness in the experiments, as this measurement was taken very close to the bed.

\begin{figure}[ht!]
	\centering
    \includegraphics[trim=0cm 0cm 0cm 0cm, clip=true, scale=1]{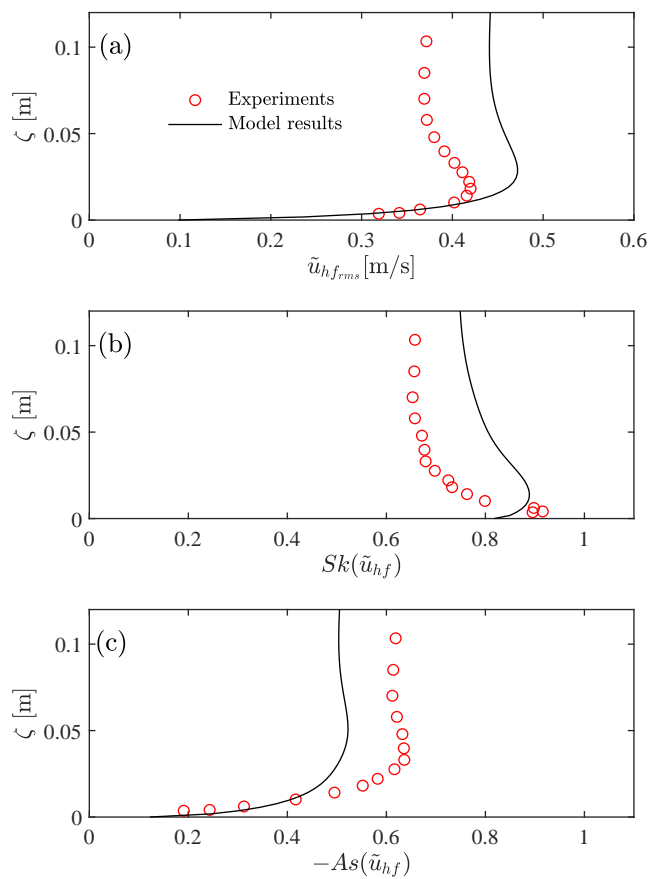}
\caption{Comparison of experimental and model results for vertical variation of (a) rms, (b) skewness and (d) asymmetry of the phase-averaged velocities, at $x=50.77$ m.}
    \label{fig:BL5077Statistics}
 \end{figure}

The present experimental conditions, as expected, generated time-averaged currents (streaming) in the boundary layer. The streaming profiles are not compared here, however, due to the model having a more pronounced bound long wave compared to the experiments (see again Figure \ref{fig:FreestreamVel}). This has been shown by \cite{Deigaardetal1999} to promote a vertical upwards as well as significant onshore shift of the streaming profile, making meaningful comparison of the streaming profiles not possible.
While it is not meaningful to compare the streaming profiles, other statistical properties, such as r.m.s., skewness and asymmetry of $\tilde u_{hf}$ can be compared. These statistical properties are important for cross-shore sediment transport and are included in many cross-shore profile models \citep[see e.g.][]{Ruessinketal2007,Dubarbieretal2015}.
Figure \ref{fig:BL5077Statistics} shows the vertical variation of the r.m.s., skewness and asymmetry of $\tilde{u}_{hf}$ of both the model and the experiments.  Figure \ref{fig:BL5077Statistics}a shows that the model generally overestimates $ \tilde{u}_{rms}$. The vertical structure is very similar and the overestimation closer to the bottom is probably a result of overestimating the free-stream value. In the experiments $\tilde{u}_{rms}$ peaks slightly closer to the bottom, indicating lower boundary layer thickness, as was also deduced from Figure \ref{fig:DetailedBL}. 
Figure  \ref{fig:BL5077Statistics}b shows that the skewness increases going from the free-stream towards the bottom. This is typical behavior of the skewness profile as shown e.g. in \cite{Bernietal2013,Henriquezetal2014,Fromantetal2019}. In the model the increase occurs further from the bed compared to the experiments and it can can be seen that the model overestimates the free-stream value while it underestimates the skewness very close to the bed. 
Finally, Figure \ref{fig:BL5077Statistics}c shows that the asymmetry decays towards the bed. This is again consistent with results from previous research \citep{Bernietal2013,Henriquezetal2014,Fromantetal2019}. The asymmetry of the model maintains a qualitatively similar vertical structure as the experiments, though it is slightly underestimated.

\subsection{Boundary layer turbulence}
\label{sec:BLk}
 In what follows we will start with a detailed comparison between the modelled and experimental boundary layer turbulence in the shoaling position, before extending the comparison and analysis to other cross-shore positions.  

\begin{figure}[ht!]
	\centering
    \includegraphics[trim=1.1cm 0.2cm 0cm 0cm, clip=true, scale=1]{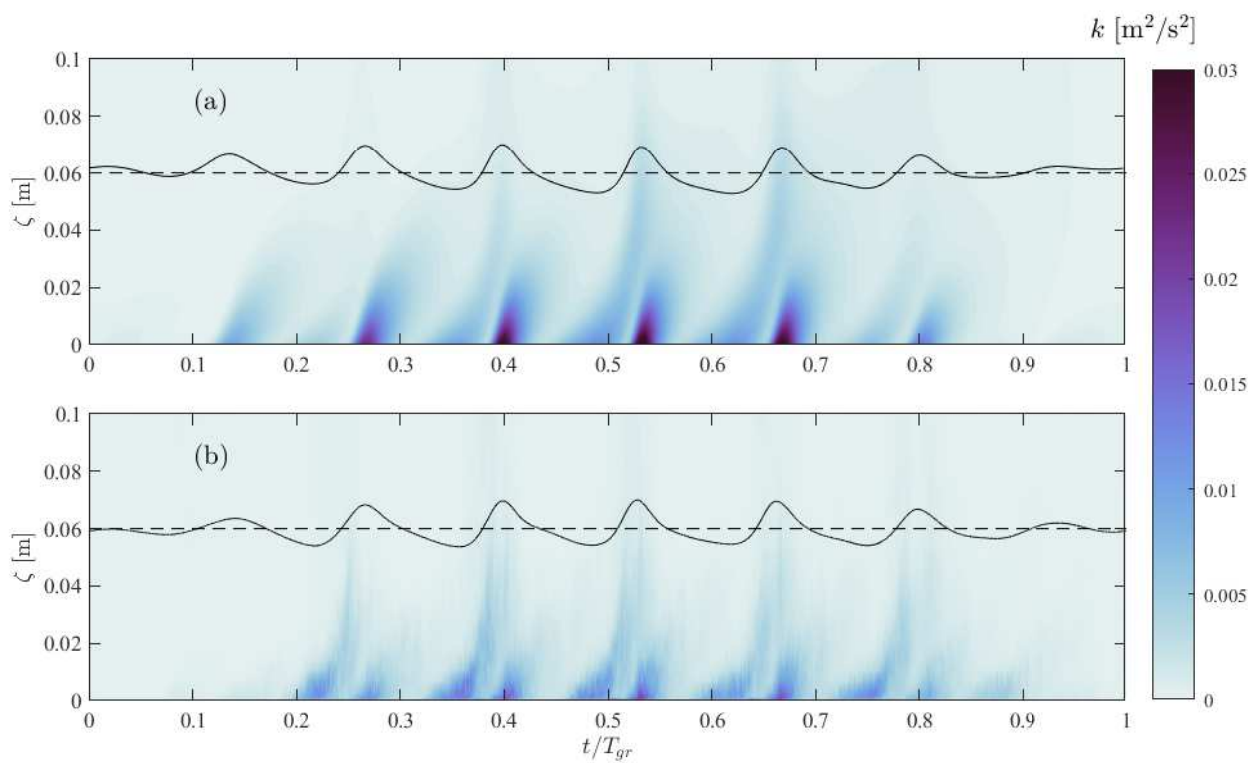}
\caption{Modelled (top) and experimental (bottom) TKE in the boundary layer at $x=50.77$m for a wave  group. The velocity at $\zeta=0.125$ m is shown as a full line as a reference.}    
    \label{fig:BL5077Turbulence}
 \end{figure}

Figure \ref{fig:BL5077Turbulence} compares the modelled and measured TKE at $x=50.77$ m as a function of the distance from the bed and phase.  In both the experiments and in the model two distinct peaks in the TKE can be seen near the bed with the passing of each individual wave. The first peak is associated with turbulence produced beneath the wave trough, and the second peak is associated with turbulence produced beneath the wave crest. The occurrence of successive peaks in near-bed TKE shows that turbulence has a high turn-over time, i.e., a major fraction of turbulent energy dissipates within one wave-cycle. As a result, any time-history effects and build-up of boundary layer TKE at the wave  group time scale are minor. This is consistent with observations of irregular oscillatory boundary layer flows in tunnels \citep{Bhawaninetal2014,YuanDash2017}. The present experimental findings suggest that time-history effects in boundary layer turbulence are also minor for full-scale progressive surface waves.
The turbulence generated at the bed beneath the wave trough is subsequently advected upwards due to vertical wave velocities and reaches higher elevations compared to the turbulence generated beneath the wave crest, when the vertical velocities during the subsequent phases are downward. This is consistent with observations under regular shoaling waves \citep{vanderZandenetal2018a} and the effect is well captured by the model. In the experiments, the TKE generated during trough and crest phases is of similar magnitude, whereas in the model, the crest-phase TKE generally exceeds the trough-phase TKE. The measured turbulence levels are subject to some uncertainty as different approaches to extract turbulence give different turbulence estimations \citep{Scottetal2005}. Therefore, it is difficult to say to which degree the model is overestimating or underestimating the turbulence beneath the crest and trough, respectively. 

The measured TKE beneath the trough is almost as high as the turbulence levels beneath the crest. Beforehand, we would have expected an image similar to the model results with higher turbulence levels beneath the crest than beneath the trough as was shown in tunnel measurements by e.g. \cite{vanderAetal2011}. The velocity skewness results in a larger onshore than offshore velocity and the acceleration skewness results in smaller boundary layer thickness during the onshore phase compared to the offshore phase. Both of these effects result in larger near-bed velocity gradients and, as result, higher expected turbulence levels beneath the crest than beneath the trough (as the production of turbulence is directly linked to the velocity gradients). The measured velocity profiles (Figure \ref{fig:DetailedBL}) show that for the experiments, the near-bed vertical gradient in velocity ($\partial u/\partial z$) is of similar magnitude during both the crest and the trough phase, which explains why the measured TKE is of similar magnitude during both stages. The modelled velocity profiles, on the other hand, show steeper velocity gradients during the crest stage, which explains why the modelled near-bed turbulence is of higher magnitude during crest than trough stage. The spatial and temporal turbulence evolution is therefore consistent with the boundary layer velocity evolution, yet the cause for the differences with the experimental results remains unclear. Additinoally, the present experimental results seem consistent with measurements of \cite{vanderZandenetal2018a} for regular progressing surface waves, which revealed highest near-bed TKE during the trough-to-crest flow reversal when large TKE production rates were observed.

\begin{figure}[ht!]
	\centering
    \includegraphics[trim=1.1cm 0.2cm 0cm 0cm, clip=true, scale=1]{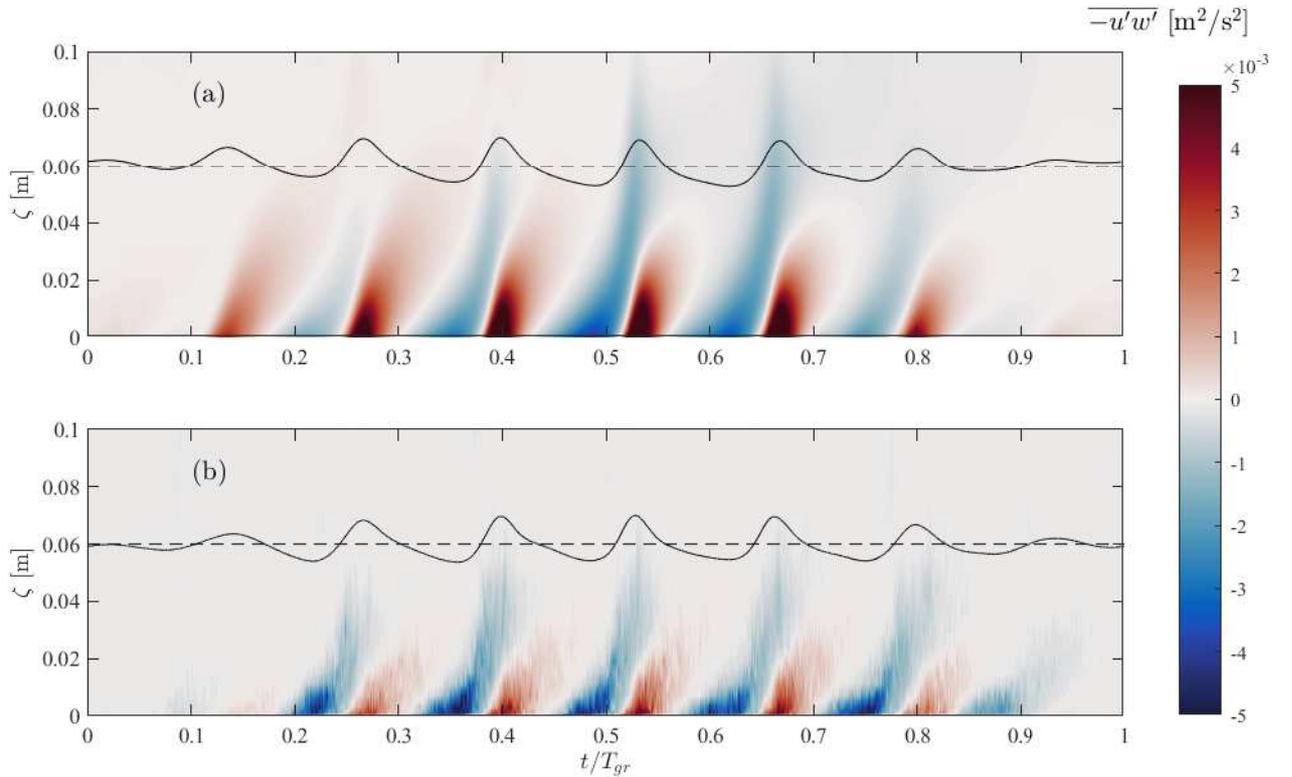}
\caption{Modelled (top) and experimental (bottom) turbulent Reynolds stress in the boundary layer at $x=50.77$m for a wave  group. The velocity at $\zeta=0.125$ m is shown as a full line as a reference.}    
    \label{fig:BL5077ReyStress}
 \end{figure}

The vertical and temporal distribution of the Reynolds stresses of both the model and the experiments (Figure \ref{fig:BL5077ReyStress}) show a similar image to that presented for the TKE (Figure \ref{fig:BL5077Turbulence}), namely that the magnitude of the Reynolds stress increases during the acceleration phases of both the positive and negative half-cycles and slightly decreases during the deceleration phases. During the offshore directed phase the negative Reynolds stresses spread to higher elevations contributing potentially to additional production of turbulence. The model captures the behaviour of the Reynolds stress reasonably well, though similar to the predictions for the TKE, the magnitude of the Reynolds stresses appear to be overestimated during the onshore phase and slightly underestimated during the offshore phase.

\begin{figure}[ht!]
	\centering
    \includegraphics[trim=0cm 0cm 0cm 0cm, clip=true, scale=1]{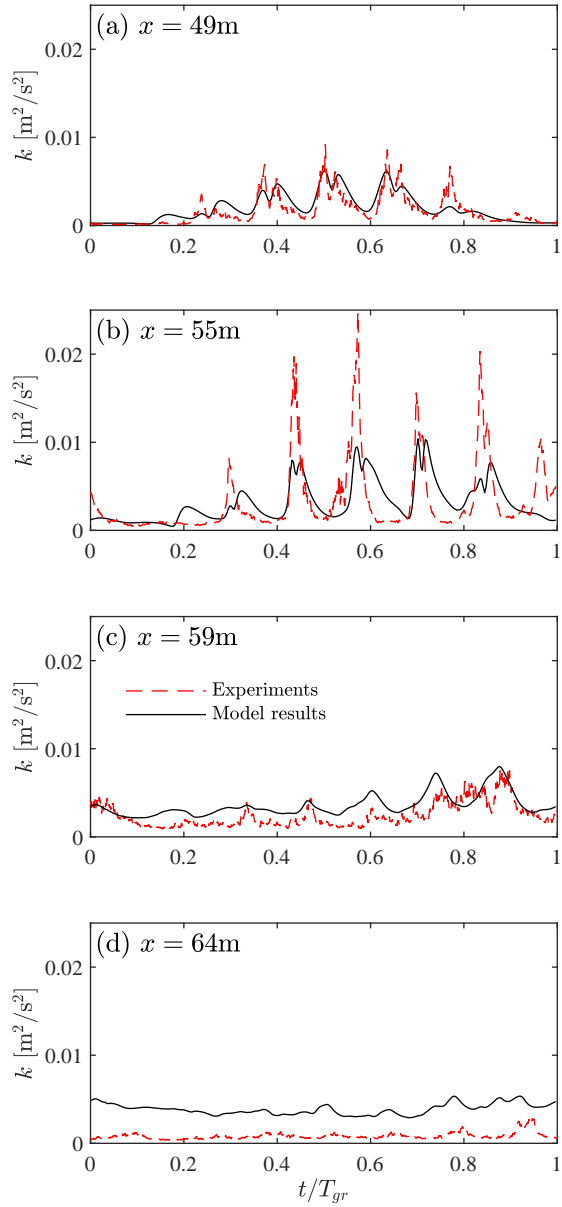}
\caption{Time series of the near-bed TKE for four cross shore positions. (a) the shoaling region, (b) at the bar crest, (c) in the bar trough and (d) in the inner surf zone.}    
    \label{fig:CrossShoreK}
 \end{figure}

Figure \ref{fig:CrossShoreK} compares the time series of the phase averaged TKE at $\zeta\approx 0.025$ m of the model and the experiment at four different cross-shore positions, corresponding to the shoaling region, the bar crest, the bar trough and the inner surf zone, respectively. 
In the shoaling position (Figure \ref{fig:CrossShoreK}a) distinct peaks in TKE can be seen. These correspond to the passing of the individual waves in the  wave group. Double peaks during successive trough- and crest-phase  cycles are observed during the passing of the three largest waves ($t/T_{gr}$=0.36--0.7). Within each double peak, the first peak is a combination turbulence from the preceding negative  cycle being advected/diffused further up into the water column and turbulence production during the upward wave zero crossing. This stage of the flow cycle is characterized by a strongly velocity-sheared flow, which, in combination with the presence of trough-phase generated turbulence, leads to a rapid increase in TKE production rates \citep{vanderZandenetal2018a}. This behaviour, as well as the general behaviour of near-bed turbulence, is well captured by the model in the shoaling position.

\begin{figure}[h!]
	\centering
    \includegraphics[trim=0cm 0.2cm 0cm 0cm, clip=true, scale=1]{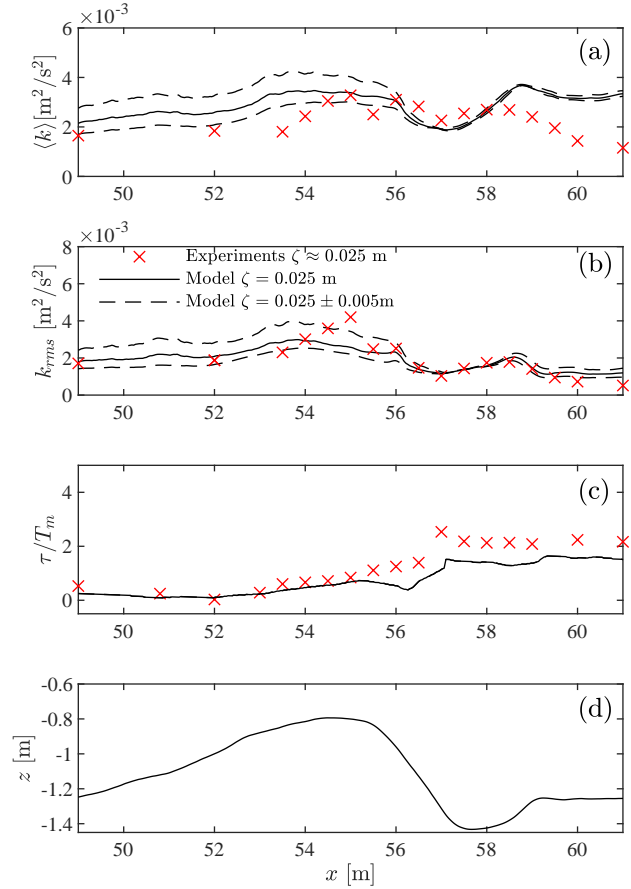}
\caption{Comparison of experimental and modelled (a) mean TKE, (b) r.m.s. of the TKE, (c) time lag of the TKE and (d) bed profile.}    
    \label{fig:kCrossNearBed}
 \end{figure}

At the bar crest (Figure \ref{fig:CrossShoreK}b) the turbulence level associated with each wave in the  wave group is significantly higher than in the shoaling position. Double peaks in TKE are still present, and the near-bed turbulence is still primarily varying with the passing of the individual waves, rather than on a wave group scale. The peaks of the TKE in the experiments are higher than in the model. This indicates that the velocity gradients are higher in the experiment than in the model or that turbulence is not transported from the bed to the same degree. This first explanation can partly be backed by Figure \ref{fig:FreestreamVel}c, where it can be seen that the peak free stream velocity is higher in the experiments than the model for three of the waves, which will lead to lower velocity gradients. It may also be related to uncertainty in defining the distance to the bed. Taking the turbulence from the model at 1.5 cm rather than 2.5 cm resulted in peaks of the same size as in the experiments (the effect of vertical position is also shown in more detail in Figure \ref{fig:kCrossNearBed}). The experiments reveal a peak in the turbulence at $t/T_{gr}=0.97$ which is absent in the model results and the model shows a small peak at $t/T_{gr}=0.2$ which is absent in the experiments. This difference is due to the modelled free-stream velocity having much smaller amplitude at $t/T_{gr}\approx 0.9$ while having a larger free-stream velocity amplitude at $t/T_{gr}\approx 0.2$ compared to the experiments (see Figure \ref{fig:FreestreamVel}c). 
Finally, it can be seen that the turbulence following the peaks decays slower in the model compared to the experiments. This may imply an underestimation of modelled turbulent dissipation rates.

Near the bar trough (Figure \ref{fig:CrossShoreK}c) the TKE peaks reach a lower level compared to those at the crest of the bar as the waves have broken and the amplitudes of the free-stream velocities are significantly smaller. In both the model and in the experiments small peaks can be seen with the passing of each individual wave, but it is also obvious that turbulence builds up gradually at the  wave group scale between $t/T_{gr} =$ 0.6--0.9. This is believed to relate to turbulence from the breaking waves penetrating into the boundary layer, consistent with the downward spreading of turbulence following each breaking wave shown in Figure \ref{fig:phaseK}. This buildup of TKE is physically explained by the large-scale breaking-generated vortices requiring multiple wave cycles to decay fully. The largest wave passes the bar crest at $t/T_{gr}=0.6$, but the turbulence in the model, as well as in the experiment, reach a maximum level at $t/T_{gr}=0.89$, thereby lagging the largest wave by two periods. This temporal lag in TKE relative to the largest wave was one of the main findings from the experiments \citep{vanderZandenetal2019} and is further discussed in relation to Figure \ref{fig:kCrossNearBed}.
In the inner surf zone  the peaks in TKE have reduced (Figure \ref{fig:CrossShoreK}d). In the experiments they are barely visible and the turbulence is generally much lower than in the model. This is consistent with the over-prediction of turbulence in the inner surf as demonstrated in Figures \ref{fig:MeanK} and \ref{fig:phaseK}.

In order to further study the near-bed TKE, 
Figure \ref{fig:kCrossNearBed} shows the cross-shore development of the time-averaged near-bed TKE (Figure \ref{fig:kCrossNearBed}a), the r.m.s. of the TKE (Figure \ref{fig:kCrossNearBed}b) and the time lag, $\tau$, of the TKE (Figure \ref{fig:kCrossNearBed}c) for both model and the experiment. Included in Figures \ref{fig:kCrossNearBed}a,b as dashed lines are also the modelled results at $\zeta=0.025 \pm 0.005$ m to show the vertical variability of these quantities. Following  \cite{vanderZandenetal2019}, the time lag was calculated by cross-correlating $k$ at the bed with the  wave group envelope at $x=50.9$ m. This particular position was chosen to prevent a bias due to the changing wave shape across the test section. The time lag $\tau$ was subsequently corrected for the changing phase of the wave  group by tracking the crest of the highest wave from $x=50.9$ m as phase reference. 

The model generally predicts the right level of TKE in the shoaling region and in the outer surf zone, whereas it clearly overestimates the mean turbulence levels in the inner surf-zone (Figure \ref{fig:kCrossNearBed}a). While maintaining approximately the right level of TKE, the cross-shore variation is not exactly captured by the model. In the experiments a fairly constant mean level was observed from $x=49$ m to $x=53.5$ m followed by a rapid increase from $x=53.5$ m to $x=55$ m, whereas the model shows a more gradual increase over the entire horizontal stretch shown in this figure. 
 A possible explanation is the uncertainty in the position of the bed in the experiments in this region. As can be seen from the model results, very large vertical variations  in mean turbulent levels occur over just one cm. This vertical variation almost disappears in the outer and inner surf zone.   

The cross-shore temporal variability in $k$, quantified by $k_{rms}$, is well captured by the model in the surf zone $x>55$ m. In the shoaling zone the experiments show a very large increase in $k_{rms}$ from from $x=53.5$ m to $x=55$ m whereas the model shows a gradual increase from $x=49$ m to $x=54$ m followed by a relatively constant level until $x=56$ m. The largest difference between the modelled and experimental  $k_{rms}$ is seen at $x=55$ m. This difference is due to the model slightly underestimating the peaks of the TKE, the model lacking a peak at $t/T_{gr}=0.97$ and finally the turbulence levels decreasing at a slower rate following each peak (Figure \ref{fig:CrossShoreK}b). The first of these three explanations may again relate to the large vertical variability in the turbulence levels. Another contributing factor may be the strongly anisotropic turbulence from the wave breaking invading the boundary layer and enhancing turbulent production as shown by \cite{vanderZandenetal2018a}. Such an effect will not be captured by a RANS model where the turbulence is isotropic. 

Finally, the model captures the time lag of the TKE well (Figure \ref{fig:kCrossNearBed}c). During shoaling the normalized time lag is generally less than 1, indicating that the turbulence locally produced with rapid turn-over and nearly in phase with the velocity. In the surf zone the normalized time lag $\tau/T_m$ is well above one, indicating that the the turbulence is externally produced (by the wave breaking) and varying at time scales larger than the individual waves. 

\section{Discussion}
\label{sec:Discussion}
This paper has focused on comparing experimental and model results for large scale bichromatic waves breaking over a breaker bar. While the model generally compares well with the experiments, there are still clear discrepancies (especially in the inner surf zone), and therefore this discussion will focus  on the model's performance in this region as well as on using 2D two-equation RANS/VOF models to simulate breaking waves. Additionally, as this paper has a special focus on processes relevant for cross-shore sediment transport, this section will address the model's applicability to simulate cross-shore sediment transport.

The results shown within this paper demonstrates that the model is able to capture surface elevations of both the shoaling and breaking waves reasonably well. This is not surprising since good surface elevation comparison with 2D RANS/VOF models has been achieved in the past, both using standard (non-stabilized) models \citep[provided that the eddy viscosity has not grown to levels leading to wave decay; see e.g.][]{LinLiu1998,Jacobsenetal2012,Brown2016} and stabilized models \citep{LarsenFuhrman2018}. The stabilized model likewise provides good accuracy for the underlying velocity kinematics in the free stream region as well as in the boundary layer. This indicates that the model would seemingly be able to predict the time varying bed shear stress which is the driver of many empirical sediment transport models. The good prediction of the wave non-linearities (skewness and asymmetry) also holds potential for improving simpler cross-shore models, as the sediment transport module inside such models often utilizes a parameterization of the wave non-linearities \citep{vanRijnetal2011a,Dubarbieretal2015}.  An existing state of the art formulation \citep{Ruessinketal2012} has improved such models significantly, but the comparison with field data shows a high degree of scatter. Recent papers \citep{Rochaetal2017,deWitetal2019} have improved on the scatter but there remains room for improvement. Specifically, the model could be used to investigate the effect of the fraction of waves breaking or the suitability of using local wave parameters on varying bathymetries, both mentioned as issues contributing to the remaining scatter \citep{deWitetal2019}.

The model is likewise able to predict the evolution in undertow structure and turbulence levels going from the shoaling region to the surf zone, something which could not be achieved using a standard (non-stabilized) turbulence model. This finding is similar to the simulation of the spilling breaking waves of \cite{TingKirby1994} presented in \cite{LarsenFuhrman2018}, and can therefore be viewed as a general performance of the stabilized model. The correct evolution in turbulence levels and undertow structure from the shoaling region through the outer surf zone is one of the major advantages of using a stabilized model opposed to a standard one.  The performance of the model in this region indicates that the model may be able to handle the suspended sediment transport in the vicinity of the bar, which is very important for both the bar position and shape. The standard model, on the other hand, with the overestimation of the turbulence levels and erroneous undertow structure, could likely overestimate the offshore directed suspended sediment transport in this region.

Because the model captures the transition in velocity profile shape it also indicates that the model could be used to study, and potentially improve, parameterizations of the spatial delay in undertow strength, which are sometimes used as a calibration parameter in simple cross-shore profile models \citep[see e.g.][]{RoelvinkStive1989,Dubarbieretal2015}. A formulation for this parameter has actually already been suggested by \cite{Jacobsenetal2014} based on their RANS/VOF simulations. 

In the inner surf zone the modelled surface elevations has a phase-lead relative to the experiments and the strength of the undertow, and TKE was overestimated by the stabilized model as well as the standard model. The discrepancies in this region are therefore not related to the stabilization of the model. This is as expected since the strain and rotation rate are of the same order of magnitude in the inner surf zone, which means that the limiter in Equation \eqref{eq:wtildeNew} will be turned off. Other 2D RANS/VOF simulations of breaking waves have likewise often resulted in overestimated undertow strengths and turbulence levels in the inner surf zone \citep[see e.g.][]{Jacobsenetal2012,Brown2016,Devolderetal2018,LarsenFuhrman2018} 
and it seems to be a common problem among many RANS/VOF simulations. There are examples of models not overestimating the undertow strength in the inner surf zone, but these have generally had a tendency to severely overestimate the turbulence levels in the outer surf zone. This is the case with the non-linear $k-\varepsilon $ model and Reynolds stress model presented in \cite{Brown2016} (see their Figures 5 and Figure 6c,d), as well as the buoyancy modified $k-\omega$ model in \cite{Devolderetal2018} (see their Figure 3b) and the stabilized model of \cite{LarsenFuhrman2018} with $\lambda_1=0$ (see their Figures 12 and 13). The overestimation of TKE in the outer surf zone with these models extracts energy from the mean flow and calms down the breaking process in the inner surf zone as a result, thereby reducing the undertow strength. Due to the severely overestimated turbulence levels, as well as the fact that these models fail to capture the velocity profile in the outer surf zone, we would argue that the good agreement in the inner surf is rather fortuitous and that these models are locally accurate for physically wrong reasons, as also suggested by \cite{LarsenFuhrman2018} in the discussion of their own results. 

In \cite{LarsenFuhrman2018} it was speculated that the exaggerated strength of the undertow in the inner surf zone was caused by an underestimation of the eddy viscosity in the upper part of the flow. Since $k$ in their case was not severely overestimated, this indicated an overestimation of $\omega$. In the present simulations $k$ is also severely overestimated in the inner surf zone (Figures \ref{fig:MeanK} and \ref{fig:phaseK}), which indicates that the problem may, in fact, not lie in an underestimation of the eddy viscosity.
It is not clear to the authors what is the core cause of the discrepancy in the inner surf zone, but we do have some thoughts on potential limitations of a 2D two-equation RANS/VOF approach. 

One limitation of this study, as well as the above mentioned studies, is that the simulations are performed in 2D, while the breaking process unquestionably is a 3D phenomenon. It might therefore be that too much of the physics are lost with the 2D simplification. One potential artifact of running a 2D simulation is that too much air can get entrained. In 3D the air will have the ability to escape from the sides of the overturning wave, but this is not possible in 2D. Whether this can explain the discrepancies in the inner surf zone is however not certain, and a comparative 2D vs. 3D study is needed to shed light on this. Additionally, in real breaking waves, the entrained air will develop into bubbles, which will subsequently rise to the surface and break up. This will not occur in the present model, where the entrained air rises as a single mixture of air and water. To account for this, either the bubbles need to be resolved, which is computationally very demanding, or a bubble model like the one presented in \cite{DerakhtiKirby2014} is needed. The inability of the model to capture the bubbles could be part of the explanation of the discrepancies in the inner surf zone, as \cite{DerakhtiKirby2014} found that the TKE in the breaking region was dampened due to the presence of the bubbles. Furthermore, they also found that the presence of bubbles increased the wave energy dissipation rate in the breaking region. Including the bubble effects presented by \cite{DerakhtiKirby2014} would thus decrease the turbulence levels in the inner surf zone, but at the same time extract more energy from the wave which could reduce the overestimation of the undertow magnitude.

In the literature several specific 3D phenoma have been reported, which the present model does not capture. These include oblique descending eddies \citep{Nadaokaetal1989}, the break up of 2D turbulent structures into 3D structures \citep{ChristensenDeigaard2001} or aereated vortex filaments \citep{LubinGlockner2015}. However, whether capturing these effects would reduce the overestimation of the undertow magnitude is not clear. In this case it is worth mentioning that performing the simulations in 3D would not necessarily mean that the model would be able to resolve all of these features. Here more computationally expensive models like LES or DNS would potentially be needed. 
Additionally, the validity of the Boussinesq approximation (linking the production of turbulence to the mean strain rate) in Equation \eqref{eq:Boussinesq} for breaking waves can be questioned. It is well known that the Boussinesq approximation fails in situation with rapidly varying strain \citep{Wilcox2006}, which is exactly what occurs when the plunging jet enters the relatively calm water in front of the wave. That the Boussinesq approximation is not a good approximation for breaking waves, can also be indicated by the relatively large importance of the $\lambda_1$ parameter on the breaking process as described in \cite{LarsenFuhrman2018} (see the difference in TKE and undertow profiles between their Cases 3, 4 and 5). The $\lambda_1$ parameter is very active in the surf zone, directly limits the production of turbulence in highly strained situations and can thereby hide the deficiencies by the Boussinesq approximation.  The $\lambda_1$ parameter is a so-called stress limiting parameter, and such stress limiters are included in many RANS models and any value between 0 and 1 is justifiable. The stress limiting feature of the $k-\omega$ SST model corresponds to setting $\lambda_1=1$ in the present model \citep[see e.g. ][for a discussion of stress limiters in RANS models]{Durbin2009}.

Finally, we would like to add that the discrepancies in the surf zone might not be caused completely by the turbulence model, but may also have to do with the solver itself. In \cite{Larsenetal2019} it was demonstrated how \verb|interFOAM| has a tendency to overestimate the crest velocities of propagating waves. A similar tendency was shown in some models by \cite{Wroniszewsketal2014}, who used four different CFD models to simulate the propagation and run-up of a solitary wave. Such an overestimation of the flow velocities in the upper part of the flow could promote the undertow shape seen in the inner surf zone, as well as the increase in TKE, as the strain rate would be overestimated.
In \cite{Larsenetal2019} it was demonstrated how the overestimation of the crest velocities could be significantly reduced for propagating waves by having a low $Co$. Whether this has sufficient effect in breaking waves is not known, however. 

The entirety of the above discussion also highlights a significant challenge in evaluating a RANS model's performance in the surf zone. Assuming that the solver performance is not optimal and that the crest velocities are overestimated \citep[as indicated in][]{Larsenetal2019}, then models might be overestimating the turbulence, not due to an erroneous turbulence model formulation, but because the strain rate is overestimated. Similarly, models producing too much turbulence could potentially perform better with regards to the undertow in the inner surf zone, as they would extract energy from the mean flow and thereby compensate for solver problems.   

Based on the discussion above, achieving better results for RANS/VOF models in the inner surf zone can be considered an important open topic for future research. It must be emphasized, however, that any modification to the models should be made while maintaining their fundamental performance e.g. in simple steady and streamwise uniform boundary layers and other canonical flows.

 \section{Conclusions}
 \label{sec:Conclusion}
 In this work a stabilized RANS model has been used to simulate large scale bichromatic waves breaking over a breaker bar. This extends previous studies using such stabilized models by focusing on plunging rather than spilling breakers, wave groups rather than regular waves, and a barred profile rather than a constant slope. Finally, novel emphasis is put on boundary layer processes, which is important for cross-shore sediment transport.

The model is compared with recent detailed measurements
and it is shown that the model handles the propagation, shoaling and breaking of the wave group.  Good agreement is achieved between modelled and measured surface elevation statistics such as the surface elevation envelope, the root mean square of the surface elevations, and wave skewness and asymmetry. Time series of the surface elevations at several cross-shore positions also compare well, and the  breaking sequence is qualitatively very similar between the model and the experiments.
 
The model predicts both mean and phase-averaged velocities and turbulence of the outer flow in the shoaling region and in the outer surf zone well. In the inner surf zone the strength of the undertow and the turbulence levels are overestimated, however. Based on previous studies using RANS models to simulate breaking waves, such discrepancies in the inner surf zone appear to be a general trend, and can be considered a topic for future research. 
Running the same simulation with a standard (non-stabilized, but still including buoyancy modification) RANS model leads to a substantial overestimation in pre-breaking turbulence levels and throughout the outer surf zone. The over-predicted turbulence levels also has an effect on the flow velocities, and it was demonstrated that the undertow (by the standard non-stabilized model) had an erroneous structure with too high offshore directed velocities in the shoaling region and in the outer surf zone. This clearly demonstrated the advantage of using a stabilized model over a standard model. 

Results from the boundary layer show that the model can handle many near bed processes that are important for cross-shore sediment transport. The model can predict the time varying boundary layer profiles and the model compares well in terms of characteristics such as velocity overshoot, boundary layer thickness, and phase lead. The model also generally reproduces the time-averaged and time-varying near-bed TKE at many cross-shore positions. Near the bed, similar to the experiments, the modelled maximum TKE in the surf zone lags the largest wave in the group by up to two wave cycles. This implies that the model captures the transport of breaking induced turbulence into the wave boundary layer. 

Overall, the performance indicates that the model would be able to improve sediment transport and morphological simulations compared to a standard (non-stabilized) model.  Furthermore, the performance of the model indicates that it  can be used to study, and potentially improve parameterizations of, processes relevant for cross-shore sediment transport, which can help improve the prediction of simpler cross-shore profile models. 

\section*{Acknowledgements}
We gratefully acknowledge the suggestions by the reviewers which helped to improve the manuscript.
The first, second and last author acknowledge support from the Independent Research Fund Denmark project SWASH: Simulating WAve Surf-zone Hydrodynamics and sea bed morphology, Grant no. 8022-00137B.
All authors additionally acknowledge support from the European Community’s Horizon 2020 Programme through the grant to the budget of the Integrated Infrastructure Initiative HYDRALAB+, Contract no. 654110, in the transnational access project HYBRID. The experimental dataset presented in this paper can be downloaded from 
https:://dx.doi.org/10.5281/zenodo.1404709

\bibliography{M:/BackUp/PHD/References/References} 

\end{document}